\documentclass[
    amsmath,
    amssymb,
    aps,
    twocolumn,
    floatfix,
    longbibliography,
    superscriptaddress
]{revtex4-1}

\usepackage[breaklinks=true, colorlinks]{hyperref}
\hypersetup{linkcolor=blue, citecolor=blue, urlcolor=blue}

\usepackage[utf8]{inputenc}
\usepackage[T1]{fontenc}
\usepackage{physics}
\usepackage{graphicx}
\usepackage{bbm, bm}
\usepackage{nicefrac}

\newcommand{\veps}{{\varepsilon}}
\newcommand{\xc}{{\text{xc}}}
\newcommand{\eff}{{\text{eff}}}
\newcommand{\exc}{{\veps _\text{xc}}}
\newcommand{\vext}{{v _\text{ext}}}
\newcommand{\rr}{{\vb{r}}}

\DeclareMathOperator*{\argmin}{argmin}

\begin{document}

\title{Neural network distillation of orbital dependent density functional theory}

\author{Matija Medvidović}
\affiliation{Institute for Theoretical Physics, ETH Zürich, CH-8093 Zürich, Switzerland}
\email{mmedvidovic@ethz.ch}

\author{Jaylyn C. Umana}
\affiliation{Center for Computational Quantum Physics, Flatiron Institute, 162 5th Avenue, New York, NY 10010, USA}
\affiliation{Department of Physics, City College of New York, New York, New York 10031, USA}
\affiliation{Department of Physics, The Graduate Center, City University of New York, New York, New York 10016, USA}

\author{Iman Ahmadabadi}
\affiliation{Joint Quantum Institute, NIST and University of Maryland, College Park, Maryland 20742, USA}
\affiliation{Center for Computational Quantum Physics, Flatiron Institute, 162 5th Avenue, New York, New York 10010, USA}
\affiliation{Department of Chemistry, Princeton University, Princeton, New Jersey 08544, USA}

\author{Domenico Di Sante}
\affiliation{Department of Physics and Astronomy, University of Bologna, 40127 Bologna, Italy}

\author{Johannes Flick}
\affiliation{Department of Physics, City College of New York, New York, New York 10031, USA}
\affiliation{Department of Physics, The Graduate Center, City University of New York, New York, New York 10016, USA}
\affiliation{Center for Computational Quantum Physics, Flatiron Institute, 162 5th Avenue, New York, New York 10010, USA}

\author{Angel Rubio}
\affiliation{Max Planck Institute for the Structure and Dynamics of Matter, Luruper Chaussee 149, 22761 Hamburg, Germany}
\affiliation{Center for Computational Quantum Physics, Flatiron Institute, 162 5th Avenue, New York, New York 10010, USA}
\affiliation{Initiative for Computational Catalysis, Flatiron Institute, 162 5th Avenue, New York, New York 10010, USA}

\date{\today}

\begin{abstract}
    Density functional theory (DFT) offers a desirable balance between quantitative accuracy and computational efficiency in practical many-electron calculations. Its central component, the exchange-correlation energy functional, has been approximated with increasing levels of complexity ranging from strictly local approximations to nonlocal and orbital-dependent expressions with many tuned parameters. In this paper, we formulate a general way of rewriting complex density functionals using deep neural networks in a way that allows for simplified computation of Kohn-Sham potentials as well as higher functional derivatives through automatic differentiation, enabling access to highly nonlinear response functions and forces. These goals are achieved by using a recently developed class of robust neural network models capable of modeling functionals, as opposed to functions, with explicitly enforced spatial symmetries. Functionals treated in this way are then called \textit{global density approximations} and can be seamlessly integrated with existing DFT workflows. Tests are performed for a dataset featuring a large variety of molecular structures and popular meta-generalized gradient approximation density functionals, where we successfully eliminate orbital dependencies coming from the kinetic energy density, and discover a high degree of transferability to a variety of physical systems. The presented framework is general and could be extended to more complex orbital and energy dependent functionals as well as refined with specialized datasets.
\end{abstract}

\maketitle

\section{Introduction}
\label{sec:introduction}

The many-electron problem has been central to quantum physics for decades. With exact solutions out of reach in most cases, different approximate methods have been used in their place, offering controlled tradeoffs between efficiency, accuracy, and applicability. Density functional theory (DFT)~\cite{hohenberg_inhomogeneous_1964, kohn_self-consistent_1965, marques_time-dependent_2006, fiolhais_primer_2003, onida_electronic_2002} should be contrasted with the accuracy of wavefunction methods such as full configuration interaction (FCI)~\cite{pople_quadratic_1987}, coupled cluster~\cite{bartlett_coupled-cluster_2007, jeziorski_coupled-cluster_1981}, quantum Monte Carlo (QMC)~\cite{becca_quantum_2017, wu_variational_2024, medvidovic_neural-network_2024, lee_twenty_2022} and Green's function methods such as dynamical mean field theory (DMFT)~\cite{georges_hubbard_1992, georges_dynamical_1996, hedin_new_1965, aryasetiawan_gw_1998, reining_gw_2018, golze_gw_2019} due to its considerably lower computational cost while still capturing the essential physics in most cases. This feature often makes DFT the only method that can access many-electron physics at large scales becoming the method of choice in solid state physics, quantum chemistry, and material science~\cite{burke_perspective_2012}.

While the existence of the exact energy functional mapping from the electron densities to energies has been proven~\cite{hohenberg_inhomogeneous_1964}, its explicit form remains unknown. Approximate forms of the exchange-correlation (XC) energy functional have been constructed at varying levels of complexity~\cite{perdew_jacobs_2001, kohn_self-consistent_1965, seidl_generalized_1996, lee_development_1988, becke_densityfunctional_1993, perdew_generalized_1996, perdew_jacobs_2001, heyd_hybrid_2003, grimme_semiempirical_2006, tkatchenko_accurate_2009, sun_strongly_2015}. More recently, expressive machine learning methods have been used to build representations of XC functionals from data~\cite{snyder_finding_2012, bogojeski_quantum_2020, bystrom_cider_2022, dick_machine_2020, lei_design_2019, chen_deepks_2021, nagai_completing_2020, m_casares_graddft_2024, del_rio_deep_2023, kirkpatrick_pushing_2021, margraf_pure_2021, bystrom_nonlocal_2024, bystrom_training_2024}, showing that achieving chemically accurate, efficient and transferable functionals is feasible, at least within a targeted domain.

In this paper, we propose a method of constructing fully machine-learned nonlocal XC functionals based on neural networks, accurately reproducing known functionals. There are two benefits of constructing a neural network copy of known functionals. First, for meta generalized gradient approximation (meta-GGA) functionals, we eliminate the orbital dependence of the kinetic contribution to the XC energy by directly controlling data generation and neural network inputs during training. Second, higher-order functional derivatives of the resulting XC functionals can easily be evaluated using automatic differentiation (AD) tools~\cite{margossian_review_2019, baydin_automatic_2018}. We call this method the global density approximation (GDA).

\begin{figure*}[!t]
    \centering
    \begin{minipage}{0.75\textwidth}
        \includegraphics[width=0.9\linewidth]{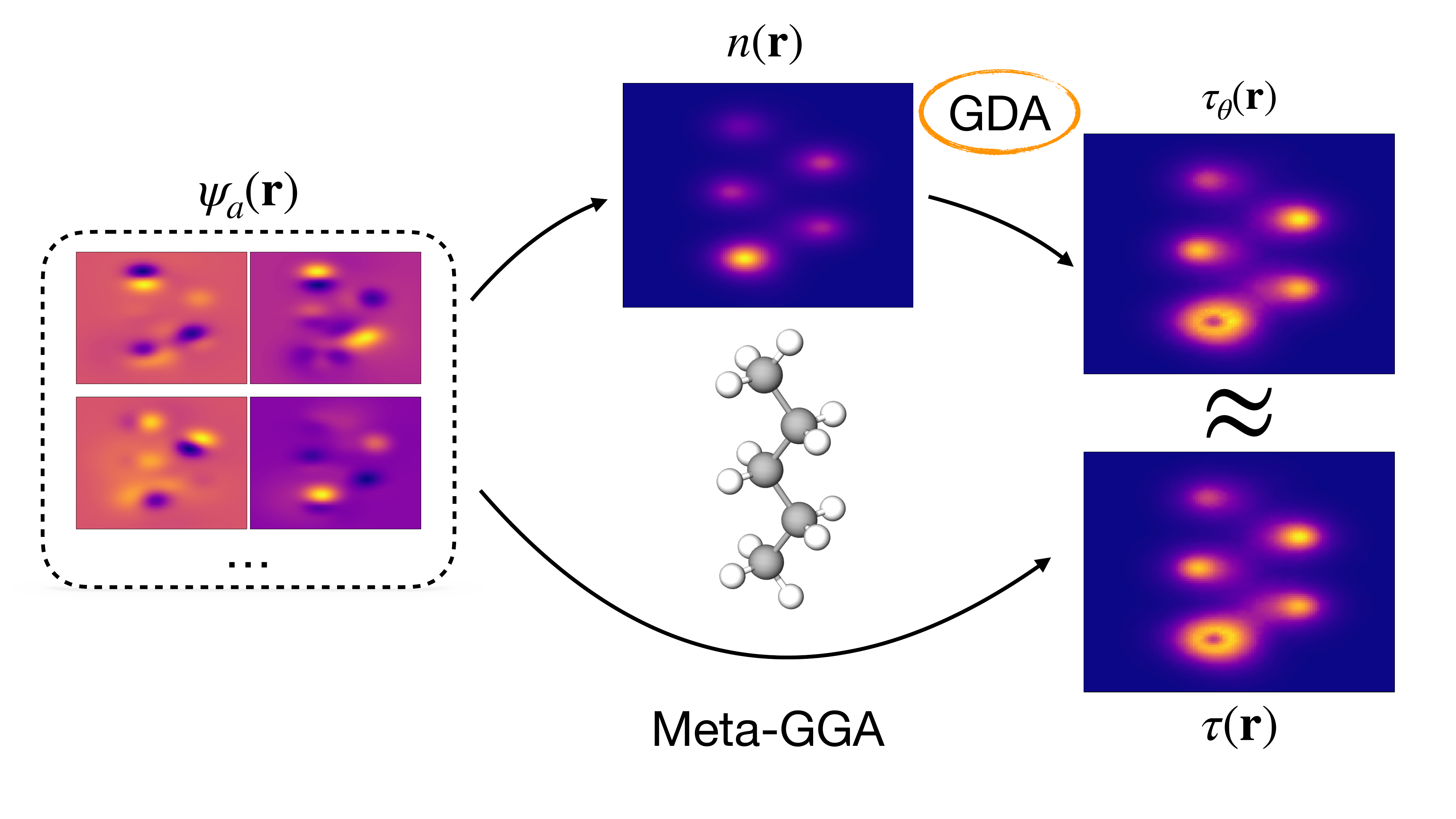}
    \end{minipage}\quad
    \begin{minipage}{0.22\textwidth}
        \includegraphics[width=\linewidth]{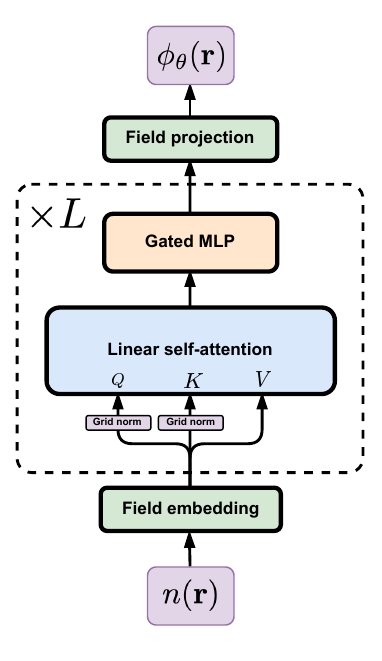}
    \end{minipage}
    \caption{
        A schematic representation of the global density approximation scheme.
        \textbf{Left}: The approximation scheme in which the kinetic energy density $\tau$ is directly inferred from the density, eliminating the orbital dependence in all resulting functionals.
        \textbf{Right}: A diagrammatic representation of the internal connectivity of the GDA model. We use $L=3$ blocks and $d=128$ as the dimension of the internal field representation.
    }
    \label{fig:diagram}
\end{figure*}

The neural network model used in this work is built around recent progress in transformer models that have recently revolutionized natural language and image processing~\cite{vaswani_attention_2017, dosovitskiy_image_2020}. We employ the linear version of the underlying attention mechanism~\cite{cao_choose_2021, li_transformer_2023, su_roformer_2023} to customize the network architecture for functional learning while respecting underlying spatial symmetries.

The GDA scheme results in an approximate but pure density functional. The main contribution of this work is the regularized training scheme allowing us to construct the GDA functional using only density-energy pairs as a part of the training dataset. As a direct result, GDA approximations are generalizable between molecules and can be used in independent self-consistent field (SCF) calculations without retraining.

\section{Methods}

\subsection{The global functional distillation}

Consider an isolated molecule with $N$ electrons in an external potential $\vext (\rr)$. We limit the discussion to isolated molecules in this paper but the discussion is equally applicable to ensembles of molecules and other quantum systems. A standard DFT calculation~\cite{martin_electronic_2020, fiolhais_primer_2003, marques_time-dependent_2006} outputs an approximation to the ground state density $n_0$ minimizing the total energy: $n_0 (\rr) = \argmin _{n} E[n]$. The total energy functional $E[n]$ is commonly written as
\begin{equation}
\label{eq:total_energy}
    E[n] = T[n] + E_\text{ext}[n] + E_H[n] + E_\xc[n]
\end{equation}
where the external contribution $E_\text{ext}[n]$ captures the effects of the external nuclear potential $\vext$ and the direct Hartree component $E_H$ is directly computable given the density $n(\rr)$. We use atomic units throughout.

A successful evaluation of Eq.~\ref{eq:total_energy} depends on approximating the unknown kinetic and exchange-correlation (XC) functionals, $T[n]$ and $E_\xc [n]$. The Kohn-Sham (KS) DFT~\cite{kohn_self-consistent_1965} framework approximates the ground state density as induced by an effective system of non-interacting electrons, $n(\rr) = \sum _a n_a | \psi _a (\rr) | ^2$, where $n_a$ is the occupation of the single-particle orbital $\psi _a (\rr)$ with energies $\epsilon _a$. Crucially, orbitals also allow for an approximate treatment of the kinetic energy contribution as $T = \int \dd[3]{\rr} \tau (\rr)$ with the kinetic energy density $\tau$ given by
\begin{equation}
\label{eq:tau}
    \tau (\rr ) = \frac{1}{2} \sum _a n_a \left| \nabla \psi _a (\rr) \right| ^2 \, .
\end{equation}
For an overview of KS-DFT, we refer the readers to the Supplemental Material and Refs.~\cite{martin_electronic_2020, jones_density_2015, kummel_orbital-dependent_2008, marques_time-dependent_2006, fiolhais_primer_2003}.

After the desired XC functional has been specified, the constrained minimization of the total energy functional given in Eq.~\ref{eq:total_energy} can proceed, yielding Kohn-Sham equations~\cite{kohn_self-consistent_1965} outlined in the Supplemental Material. These equations have to be solved in a self-consistent (SCF) manner, ensuring that $n(\rr) = \sum _a n_a | \psi _a (\rr) | ^2$ holds at all times.

The exact energy functional is unknown. However, approximations with increasing levels of complexity have been ordered into the so-called \textit{Jacob’s ladder}~\cite{perdew_jacobs_2001}. Higher rungs capture more details of local density neighborhoods at increased computational cost, starting with the local density approximation (LDA) which approximates electron XC effects as a uniform gas of interacting electrons of density $n (\rr)$~\cite{perdew_self-interaction_1981}. While providing a good approximation for systems with slowly varying densities, it often lacks accuracy for molecular systems. This motivated the development of the second rung, the generalized gradient approximation (GGA)~\cite{langreth1983beyond, perdew_generalized_1996}, which includes local gradients $\nabla n (\rr)$ and $\nabla ^2 n(\rr)$ as a local variable, offering better accuracy for a wider range of chemical systems.

\begin{figure*}[!t]
    \centering
    \begin{minipage}{0.32\textwidth}
        \includegraphics[width=\linewidth]{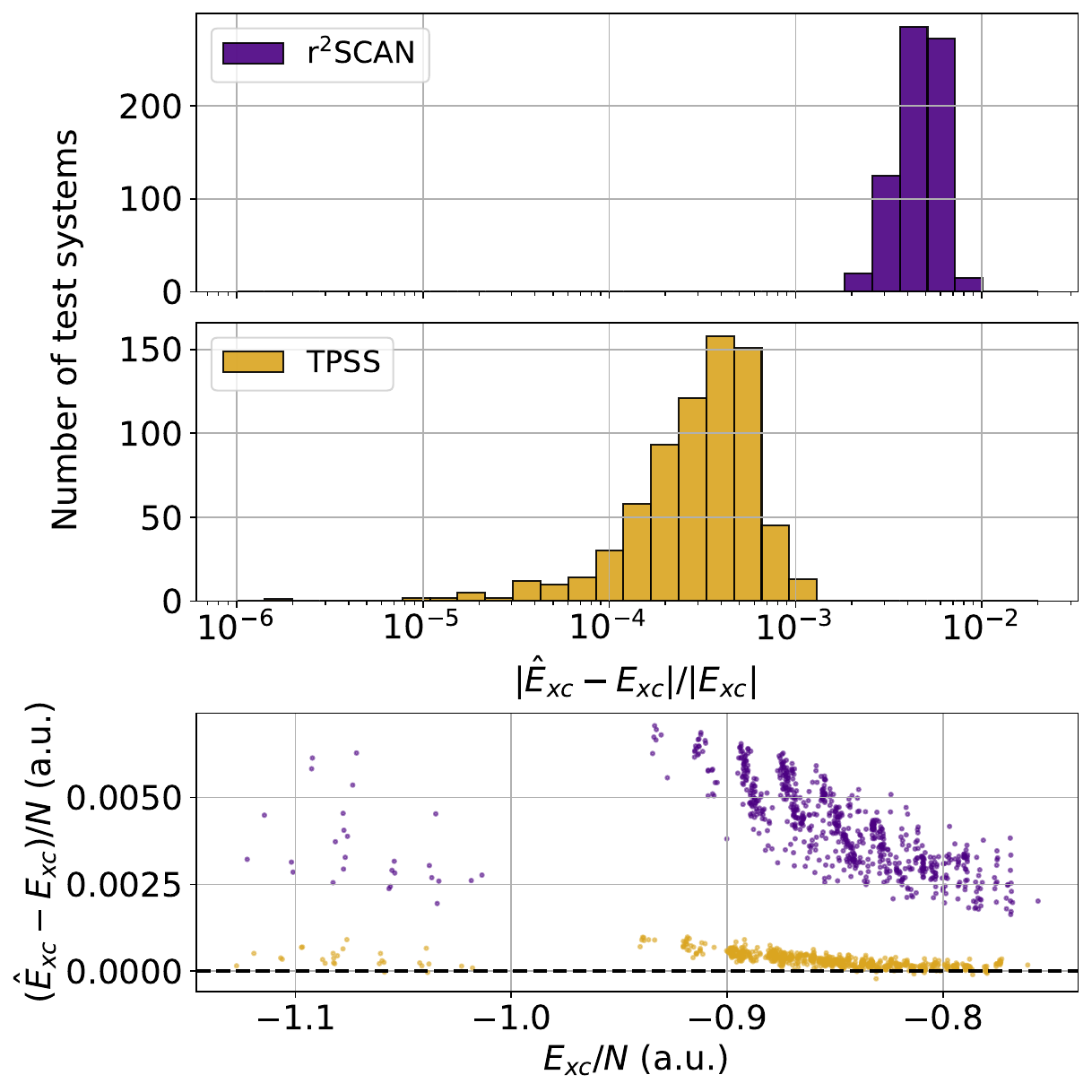}
    \end{minipage}\quad
    \begin{minipage}{0.32\textwidth}
        \includegraphics[width=\linewidth]{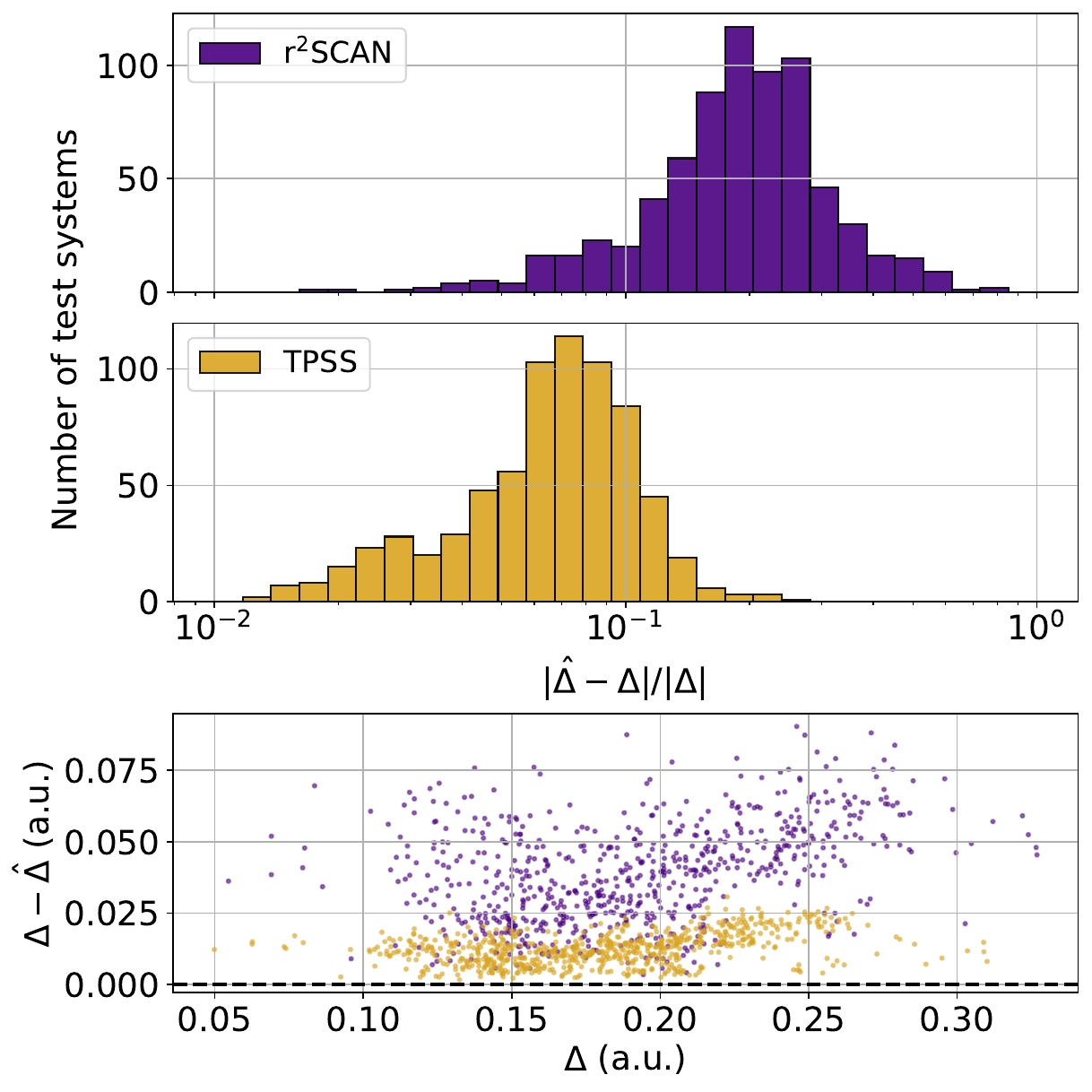}
    \end{minipage}\quad
    \begin{minipage}{0.32\textwidth}
        \includegraphics[width=\linewidth]{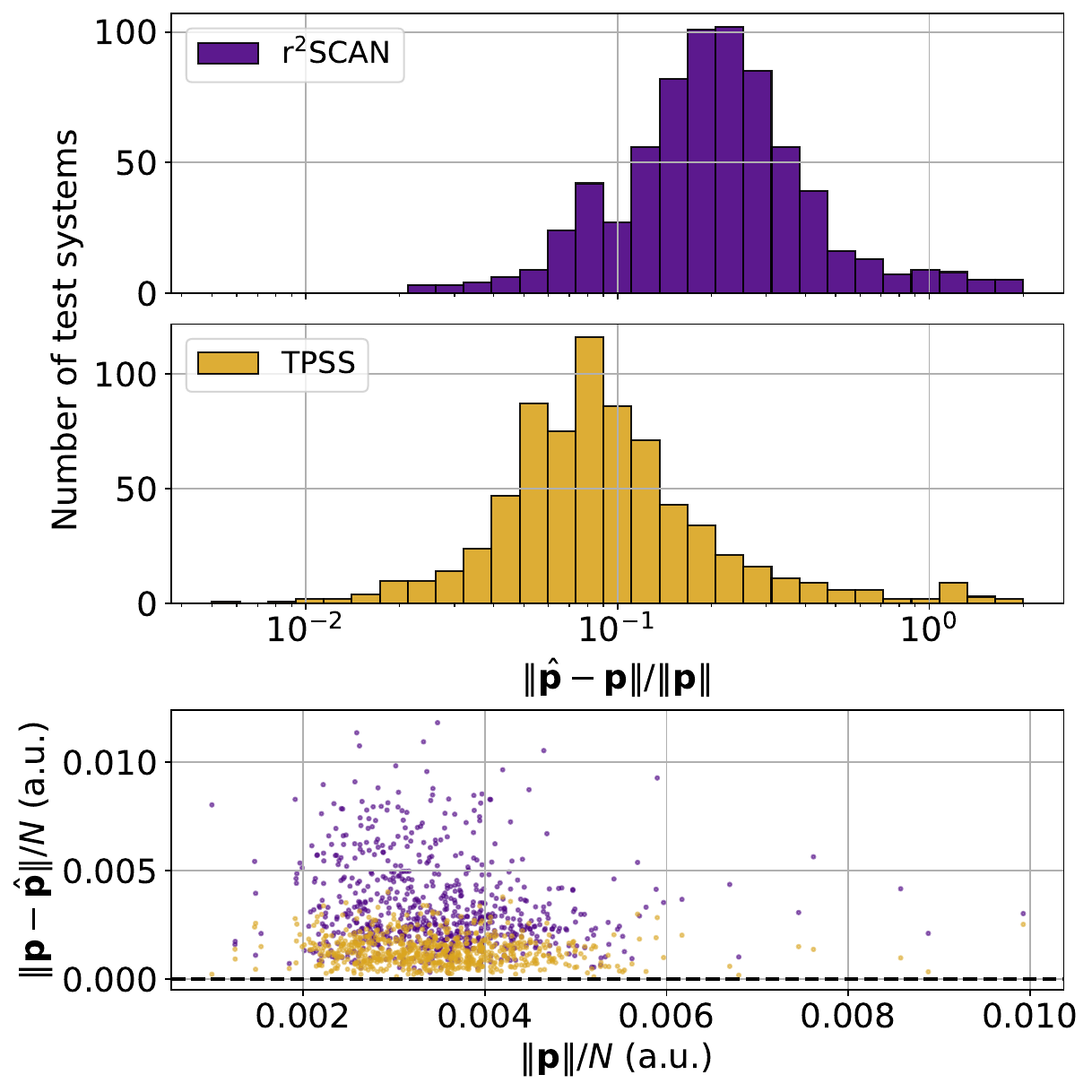}
    \end{minipage}
    \caption{
        Error distributions of observables obtained from first-principles calculations using the GDA approximation. The surrogate functional was tested on 717 test molecular systems withheld during training, a random selection of $10 \%$ of the QM7~\cite{blum_970_2009, rupp_fast_2012} dataset.
        \textbf{Left}: XC energy values.
        \textbf{Middle}: HOMO-LUMO gap values.
        \textbf{Right}: Total dipole moments.
    }
    \label{fig:observables}
\end{figure*}

Even higher rungs host the meta-GGA functionals~\cite{tao_climbing_2003, perdew_workhorse_2009, zhao_new_2006, sun_effect_2012, sun_semilocal_2013, sun_strongly_2015, sun_accurate_2016, bartok_regularized_2019, furness_accurate_2020, becke_simple_2006, tran_accurate_2009}, which capture more complex electronic interactions and offer improved accuracy for diverse systems, especially in terms of chemical reactivity and band gaps. These benefits come from including new local variables such as $\tau (\rr )$ and its orbital dependence via Eq.~\ref{eq:tau}.

Mathematically, the capacity to numerically solve the KS equations and access observables relies on our ability to approximate functional derivatives of the total energy in Eq.~\ref{eq:total_energy} during the SCF loop. In this paper, we propose an approximation to efficiently estimate functional derivatives of common meta-GGA density functionals by systematically removing orbital dependence. We do this by fitting them to expressive parameterized neural network models with restricted input variables. After capturing the target functional to a satisfactory degree, derivatives of the model can be efficiently calculated using automatic differentiation tools~\cite{margossian_review_2019, baydin_automatic_2018, Paszke2017}.

We rewrite the kinetic energy density as $\tau \approx \tau _\theta [n]$, a nonlocal functional captured by an expressive deep neural network~\cite{lecun_deep_2015, goodfellow_deep_2016, dawid_modern_2022} with $\theta$ indicating a set of all free parameters in neural network sub-components (see Supplemental Material for details). This approximation allows us to trivially rewrite any meta-GGA functional as
\begin{equation}
\label{eq:gda}
    E ^{\text{GDA}} _\xc [n] = \int \dd[3]{\rr} \, n (\rr) \, \exc ( n(\rr), \nabla n (\rr), \tau _\theta [n] (\rr) ) \, ,
\end{equation}
turning it into a nonlocal density functional.

\begin{figure*}[!t]
    \centering
    \begin{minipage}{0.49\textwidth}
        \includegraphics[width=\linewidth]{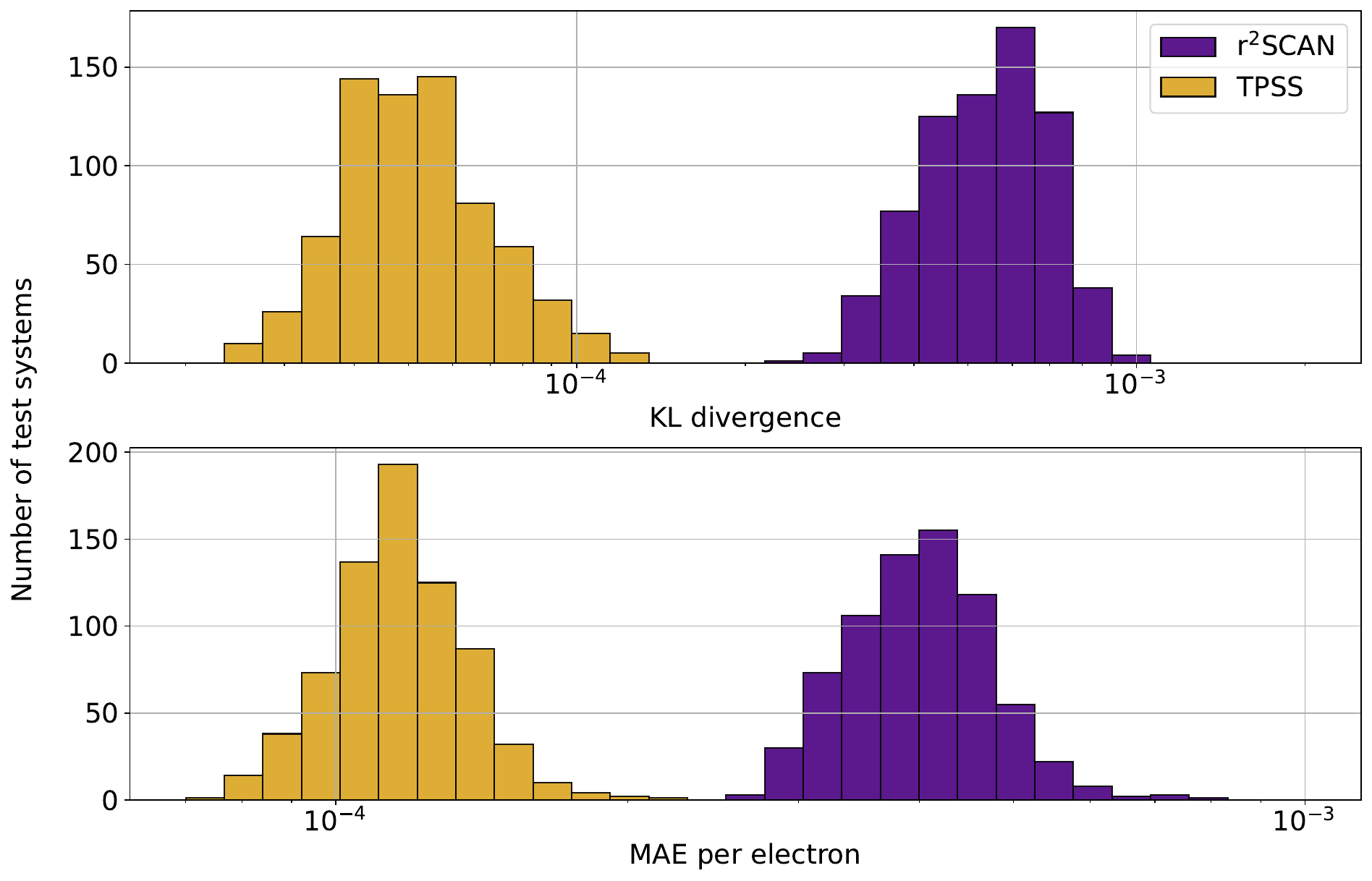}
    \end{minipage}\quad
    \begin{minipage}{0.49\textwidth}
        \includegraphics[width=0.95\linewidth]{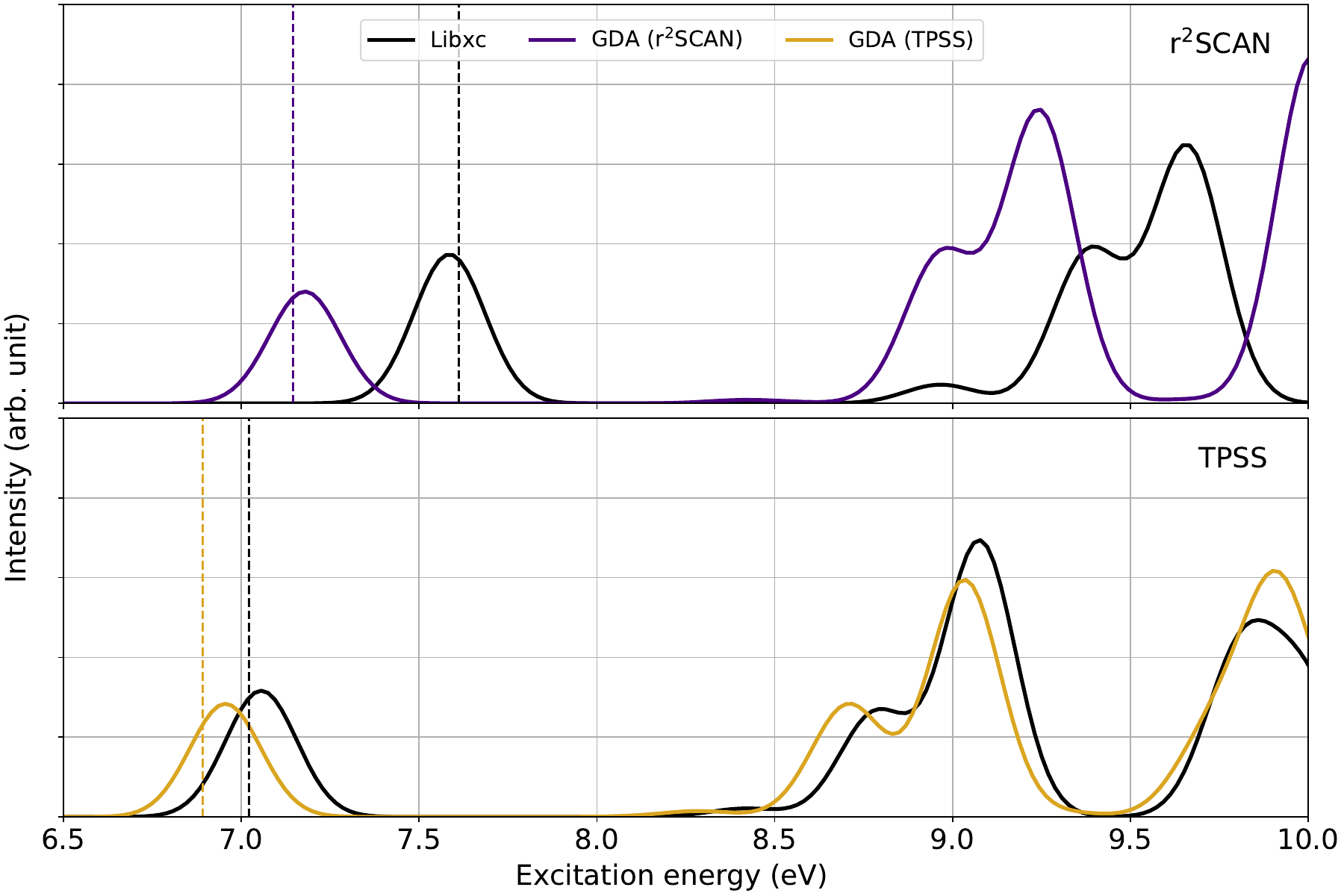}
    \end{minipage}
    \caption{
        Comparison of first-principles densities calculated using independent KS-DFT calculations and an example linear-response calculation of dimethyl ether (CH$_3$OCH$_3$) spectra using TD-DFT where the predicted KS gaps are indicated by vertical lines in the same style as the corresponding spectra. We note that the spectral data are shown as a proof-of-principle, as an example of a calculation that can be performed with more specialized GDA functionals using automatic differentiation.
    }
    \label{fig:density-spectrum}
\end{figure*}

\subsection{The neural network model}
\label{sec:ml_details}

In Eq.~\ref{eq:gda}, we parametrize the kinetic energy density $\tau _\theta [n]$ as a deep neural network, capable of approximating functionals in a controlled way while respecting the underlying spatial symmetries. Additionally, to satisfy known exact scaling laws, we model the local dimensionless enhancement factor with a parametrized form $\phi_\theta [n] (\rr)$, such that
\begin{equation}
\label{eq:phi-def}
    \tau _\theta (\rr) = \tau _W(\rr) + e^{\phi _\theta (\rr) } \left( \tau_U (\rr) + \eta \, \tau_W (\rr) \right)
\end{equation}
is represented by a neural network with parameters $\theta \in \mathbbm{R} ^P$, based on a dimensionless kinetic energy indicator variable used in Ref.~\cite{furness_accurate_2020}. In Eq.~\ref{eq:phi-def}, $\tau_U = \frac{3}{10} (3 \pi ^2 )^{\nicefrac{2}{3}} n ^{\nicefrac{5}{3}}$ is the uniform electron gas kinetic energy, and $\tau_W = \flatfrac{| \nabla n |^2}{8 n}$ is the von Weizsäcker~\cite{weizsacker_zur_1935} kinetic functional. The regularization parameter was set to $\eta = 10^{-3}$, as defined in Ref.~\cite{furness_accurate_2020} and informed by normalization heuristics~\cite{subramanian_towards_2023, mccabe_multiple_2023} based on exact values of $\tau (\rr)$ in the dataset.

Parameterizing only dimensionless enhancement factors in Eq.~\ref{eq:phi-def} allows us to recover the correct scaling laws for the output $\tau _\theta (\rr)$. Therefore, the GDA approximation inherits all scaling properties from parent meta-GGA functionals, such as the uniform scaling of the exchange functional.

Our architecture is similar to an encoder-only transformer~\cite{vaswani_attention_2017} with linear attention modules~\cite{li_transformer_2023, cao_choose_2021}. Attention layers were chosen because of the permutation equivariance property -- permuting input values on the grid also permutes outputs in the same way, making this architecture well suited for functional learning on grids.

We highlight two main differences in internal component design: a flexible density embedding layer, lifting the density into a high-dimensional representation while respecting underlying symmetries, and specialized linear attention layers to directly operate on DFT grids equipped with custom normalization sub-components.

Local molecular densities can vary several orders of magnitude, depending on proximity to nuclei. To normalize density variations, we construct the following dimensionless $d$-component input field:
\begin{equation}
\label{eq:phi}
    \phi (\rr) = \overline{\ln n(\rr)} \, \vb{e}_n + \overline{\ln | \nabla n(\rr) | ^2} \, \vb{e}_\gamma \; ,
\end{equation}
where $\vb{e} _n \, , \vb{e} _\gamma \in \mathbbm{R}^d$ are trainable vectors and $\overline{f(\rr)}$ indicates subtracting the mean and dividing by the standard deviation of $f(\rr)$ with respect to the distribution $n(\rr)/N$. Input $\phi$ values are updated by subsequent network layers.

To make $\phi _\theta$ invariant with respect to translations and rotations of input coordinates we use a symmetry-aware \textit{positional encoding} sub-layer. All computations are performed in a coordinate system where the mean value (dipole moment) of the density distribution $n(\rr)/N$ vanishes and the covariance matrix is diagonal. This choice eliminates the special Euclidean symmetries in $\text{SE(3)}$, determining the resulting computational coordinate values up to molecular point-group symmetries.

The coordinates are then lifted to a $d$-dimensional representation using random Fourier features (RFFs)~\cite{rahimi_random_2007, tancik_fourier_2020, li_learnable_2021} combined with the density embedding of Eq.~\ref{eq:phi} using a gating mechanism described in Ref.~\cite{shazeer_glu_2020}. The lifted density representation is then processed by a sequence of $L$ blocks. One GDA block is defined as a stack of self-attention (SA) and gated multi-layer perceptron (MLP) layers~\cite{shazeer_glu_2020} with $\texttt{SiLU}$ activations~\cite{elfwing_sigmoid-weighted_2017}. After $L$ blocks, one final projection layer consisting of an element-wise gated MLP is applied, projecting the point-wise embeddings into a single real number per coordinate $\rr$, which we then interpret as the final value of the field $\phi = \phi _\theta [n] (\rr)$ in Eq.~\ref{eq:phi-def}. We refer the reader to Fig.~\ref{fig:diagram} for an overview of the internal connectivity and the Supplemental Material for numerical and technical details.

The nonlocality in the GDA model is captured by the linear attention layer~\cite{cao_choose_2021, li_transformer_2023}. The input field $\phi$ is transformed as:
\begin{equation}
\label{eq:linear-attn}
    \phi ' (\rr) = \int \dd[3]{\rr'} n(\rr') \left( Q (\rr) \cdot K(\rr') \right) V(\rr')
\end{equation}
where $Q_i(\rr), K_i(\rr), V_i(\rr)$ are the \textit{query}, \textit{key} and \textit{value} parametrized local projections of the input field: $ \sum _k W^{Q, K, V} _{i k} \phi _k (\rr)$. Furthermore, we employ rotary positional encoding (RoPE)~\cite{su_roformer_2023, li_transformer_2023} independently for each block to ensure that the product $Q(\rr) \cdot K(\rr') = \mathcal{K}(\rr - \rr ')$ depends only on the relative coordinate, parametrizing a flexible continuous convolutional kernel.

Apart from facilitating translational invariance, we emphasize that the linear attention mechanism used in this work sidesteps the unfavorable quadratic scaling of simple functional evaluations. As a consequence, using the transformer architecture does not spoil asymptotic scaling properties of the SCF loop. Further details about the model used in this work are given in the Supplemental Material

The GDA model $\phi _\theta$ is trained using gradient-based optimization of parameters $\theta$ using the RAdam optimizer~\cite{liu_variance_2019, kingma_adam:_2015, loshchilov_fixing_2017} and the gradients~\cite{paszke_pytorch_2019} of the scalar cost function
\begin{equation}
\label{eq:cost}
    \mathcal{C} (\theta) = \frac{\norm{\phi _\theta - \phi _0} _G ^2}{\norm{\phi _0} _G ^2 } + \lambda \frac{\norm{\mathcal{T} - \mathcal{T}_0} _F ^2}{ \norm{\mathcal{T}_0} _F ^2}
\end{equation}
consisting of two terms. The first term is the scaled mean squared error (MSE) for $\phi$ itself where the unweighted grid norm $\norm{f} _G ^2 = \sum _i f(\rr _i) ^2$ is defined over all DFT grid points $\{\rr _1, \ldots , \rr _{N_g} \}$, where $\phi _0 (\rr)$ is evaluated from precalculated $\tau$ values in the dataset, by inverting Eq.~\ref{eq:phi-def}. The second term in Eq.~\ref{eq:cost} enforces the correct predicted kinetic energy matrix $\mathcal{T}$ in the KS orbital basis, which can be easily predicted from the one-body electronic density matrix $\Gamma$ as
\begin{equation}
\label{eq:derivative-loss}
    \mathcal{T} _{a b} = \sum _{\mu \nu} C_{\mu a} C_{\nu b} \, \pdv{T[n]}{\Gamma _{\mu \nu}}
\end{equation}
by using standard AD routines and the linear combination of atomic orbitals (LCAO) expansion coefficients $C_{\mu a}$, see Supplemental Material Reference values $\mathcal{T} _0$ can be precalculated as $\mathcal{T} _{0, ab} = \frac{1}{2} \int \dd[3]{\rr} \nabla \psi _a \cdot \nabla \psi _b $ from easily accessible basis set integrals~\cite{sun_pyscf_2018, sun_recent_2020}.

Including the second term in the cost function in Eq.~\ref{eq:cost} ensures that the kinetic contribution to the overall KS effective Hamiltonian $H _\eff = \nicefrac{\partial E}{\partial \Gamma}$ is well approximated by the GDA surrogate functional when solving the KS equation $C^\top H _\eff \, C = \epsilon $. We find that including such a gradient cost term is key for making the resulting neural network functional converge in a practical DFT SCF calculation, allowing us to use it to obtain energies, orbitals, densities or observables from first principles, without ever referring to the parent functional. Similar regularization methods have been proposed in Refs.~\cite{mazo-sevillano_variational_2023, kirkpatrick_pushing_2021}.

\section{Results}
\label{sec:results}

We examine GDA approximations to two prominent meta-GGA functionals: r${}^2$SCAN~\cite{furness_accurate_2020, sun_strongly_2015, sun_accurate_2016} and TPSS~\cite{tao_climbing_2003}. We use a single neural-network GDA model for evaluation of all molecules for all three functionals. Each functional is tested in both ground state KS-DFT and linear response time-dependent density functional theory (TD-DFT) calculations. The architecture outlined in Fig.~\ref{fig:diagram} with $L=3$ blocks and the embedding dimension of $d=128$ is used and optimized using $\lambda = 1$ in Eq.~\ref{eq:cost}. Molecular geometries from the QM7~\cite{blum_970_2009, rupp_fast_2012} dataset are used, consisting of 7165 organic molecules of up to 23 atoms, and featuring a large variety of molecular structures such as double and triple bonds, cycles, carboxy, cyanide, amide, alcohol and epoxy. Corresponding KS-DFT calculations were done using the r${}^2$SCAN functional to obtain input densities and target $\tau$ values.

We independently calculate and compare several physical observables: the total energy, the molecular dipole moment, the KS highest occupied molecular orbital – lowest unoccupied molecular orbital (HOMO-LUMO) gap, and the self-consistent density itself. Error distributions are shown in Fig.~\ref{fig:observables}, demonstrating that the resulting surrogate GGA functionals converge to physical results independently from the source meta-GGA functional, using only density inputs as shown in Eq.~\ref{eq:gda}.

The GDA functional is able to accurately predict XC energies and potentials over a large range of diverse test molecules, eliminating orbital dependence from input functionals. Gap values are predicted within $~10\%$ for TPSS with higher errors in r${}^2$SCAN calculations where GDA approximations tend to over estimate the gap. A similar trend persists with dipole moments, demonstrating qualitative accuracy. Since all of the results have been obtained with a single transferable neural network model, the GDA approximation scheme eliminates the need to train multiple models for de-orbitalizing different XC functionals and can serve as a starting point for more fine-tuned approximations. Furthermore, we speculate that such broad generalization for larger models trained on more diverse datasets can be used to train foundation functionals to then fine tune on downstream problems with limited data.

As a direct global divergence measure between two densities, we consider the Kullback-Liebler (KL) divergence~\cite{kullback_information_1951} $ D_\text{KL} (n \parallel \Hat{n}) = \frac{1}{N} \int \dd[3]{\rr} \, n(\rr) \, \ln \frac{n (\rr)}{\Hat{n} (\rr)}$ commonly used in statistical literature, where $\Hat{n}$ is the density obtained by independent SCF convergence using the GDA approximation of the original functional. In addition, we also compare direct mean absolute errors (MAEs) of density values defined as $D_\text{MAE} (n, \Hat{n}) = \frac{1}{N} \int \dd[3]{\rr} \, \left| n(\rr) - \Hat{n}(\rr) \right| $. In all cases the trained model produces a good approximation to the ground-state density indicating good transferability, as can be seen in the left panel of Fig.~\ref{fig:density-spectrum}. We see that TPSS densities are reproduced more accurately indicating that the generated dataset based on QM7 is better suited to some typical densities encountered with that functional. A more expansive dataset with more diverse densities and geometries is likely to close the gap in Fig.~\ref{fig:density-spectrum}.

To showcase the performance of GDA approximations in the excited state regime, we present a qualitative demonstration in the right panel of Fig.~\ref{fig:density-spectrum}. Here, we simulate absorption spectra via linear response for a single test molecule. We compare spectra generated using the GDA model with the parent functionals accessed through the LibXC~\cite{marques_libxc_2012, lehtola_recent_2018} library. Calculations are carried out in the Casida TD-DFT~\cite{casida_time-dependent_1995} formalism for 100 excited states, and we focus on a low energy range to highlight the ability of our model to replicate physically meaningful features before the ionization threshold~\cite{tam_electron_1974, linstrom_nist_2001} in a fashion similar to the original functional. Unsurprisingly, the GDA scheme captures the behavior of TPSS. However, for most test molecules in our dataset, the excited state energy values are qualitative. For example, in the case of r${}^2$SCAN, we observe spectral shifts (on the order of $\sim 1$ eV in Fig.~\ref{fig:density-spectrum}) but still capture the overall shape and widths of the affected features. The GDA model has a higher error in r$^2$SCAN with the expectation that the predictions can be improved by using larger GDA neural networks trained on more specialized datasets. Therefore, the spectral data is shown as a proof of principle, as an example of a linear response calculation that can be performed with fine-tuned GDA functionals using automatic differentiation.

\section{Conclusion and outlook}
\label{sec:conclusion}

We introduce another class of approximations, enabling first-principles replacement of orbital-dependent meta-GGA functionals. Using these approximations, arbitrary derivatives of source functionals can be constructed, enabling access to different results and phenomena (e.g. highly nonlinear responses).

As a proof of principle, this approximation is accurate and resource-efficient with a high degree of transferability between different functionals and molecular systems. Tests on periodic systems are left for future research. In addition, our GDA approximation scheme formally enables the use of orbital-free DFT (OF-DFT) calculations at the meta-GGA level of theory. However robust OF-DFT solvers for Gaussian-type basis sets are still an active area of research with experimental support for nonlocal functionals at best. Therefore, extending the GDA approximation to OF-DFT is left for future research.

More general orbital-dependent functionals have shown great potential in overcoming the limitations of their pure density functionals. Hybrid functionals, such as Becke three-parameter Lee-Yang-Parr (B3LYP), Heyd-Scuseria-Ernzerhof (HSE) and Perdew-Burke-Ernzerhof 0 (PBE0)~\cite{stephens_ab_1994, heyd_hybrid_2003, adamo_toward_1999}, which combine GGA functionals with a fraction of orbital-dependent exact exchange, have achieved superior accuracy in predicting molecular geometries, reaction barriers, and electronic properties. The GDA treatment of the Fock operator is a topic of ongoing research. The GDA approach allows reformulating orbital-dependent functionals as pure but nonlocal density functionals, offering a promising direction to bridge the gap between \textit{pure} DFT and the quantum chemical accuracy.

\subsection*{Software and simulations}

All DFT simulations were performed using a custom interface between the PySCF~\cite{sun_pyscf_2018, sun_recent_2020} library and PyTorch~\cite{Paszke2017, paszke_pytorch_2019}, used for automatic generation of XC potentials for KS-DFT calculations and kernels for linear-response TD-DFT calculations.

All calculations were performed using the correlation-consistent polarized valence double zeta (cc-pVDZ) basis set at grid level 1 in PySCF. Convergence tolerance was set to $10^{-6}$ Ha. Code needed to reproduce results in this work or experiment with new results has been open-sourced and can be found at GitHub: \url{https://github.com/Matematija/global-density-approximation}.

\subsection*{Acknowledgements}

M.M. acknowledges support from the CCQ Graduate Fellowship in computational quantum physics under the Grant No. 653217. J.C.U. acknowledges support from the CCQ Graduate Fellowship in computational quantum physics under the Grant No. 1165064.

\bibliography{references}

\end{document}


\title{
    Neural network distillation of orbital dependent density functional theory\\
    \vspace{3ex}
    Supplementary Material
}

\author{Matija Medvidović}
\affiliation{Institute for Theoretical Physics, ETH Zürich, CH-8093 Zürich, Switzerland}

\author{Jaylyn C. Umana}
\affiliation{Center for Computational Quantum Physics, Flatiron Institute, 162 5th Avenue, New York, NY 10010, USA}
\affiliation{Department of Physics, City College of New York, New York, NY 10031, USA}
\affiliation{Department of Physics, The Graduate Center, City University of New York, New York, NY 10016, USA}

\author{Iman Ahmadabadi}
\affiliation{Joint Quantum Institute, NIST and University of Maryland, College Park, MD 20742, USA}
\affiliation{Center for Computational Quantum Physics, Flatiron Institute, 162 5th Avenue, New York, NY 10010, USA}
\affiliation{Department of Chemistry, Princeton University, Princeton, NJ 08544, USA}

\author{Domenico Di Sante}
\affiliation{Department of Physics and Astronomy, University of Bologna, 40127 Bologna, Italy}

\author{Johannes Flick}
\affiliation{Department of Physics, City College of New York, New York, NY 10031, USA}
\affiliation{Department of Physics, The Graduate Center, City University of New York, New York, NY 10016, USA}
\affiliation{Center for Computational Quantum Physics, Flatiron Institute, 162 5th Avenue, New York, NY 10010, USA}

\author{Angel Rubio}
\affiliation{Max Planck Institute for the Structure and Dynamics of Matter, Luruper Chaussee 149, 22761 Hamburg, Germany}
\affiliation{Center for Computational Quantum Physics, Flatiron Institute, 162 5th Avenue, New York, NY 10010, USA}
\affiliation{Initiative for Computational Catalysis, Flatiron Institute, 162 5th Avenue, New York, NY 10010, USA}

\date{\today}

\maketitle

\section{Neural network architecture}

\subsection{Linear attention layers as continuous convolutions}
\label{appendix:attention}

We consider input data in a point cloud format of density values with coordinates $\rr _i$, density values $n _i = n (\rr _i)$ and quadrature weights $w_i$ where $i=1, \ldots, N_g$ indexes the cloud in arbitrary order. Functional inputs in this form are general and can be readily extracted from modern quantum chemistry software packages. This enables us to approximate integrals of sufficiently well-behaved functions as:
\begin{equation}
    \int \dd[3]{\rr} f(\rr) \approx \sum _i w_i \, f(\rr _i) \; .
\end{equation}

Global density approximations rely on the linear attention~\cite{cao_choose_2021} mechanism for propagating information between local density values. In purely mathematical terms, the linear self-attention operation used in this work can be seen as a discretization of an integral transform with a learnable kernel $\mathcal{K}$:
\begin{equation}
\label{eq:linear-attn-def}
    \phi ' (\rr) = \int \dd[3]{\rr '} n(\rr ') \mathcal{K}(\rr, \rr ') \phi (\rr ') =
    \int \dd[3]{\rr '} n(\rr') \left( Q (\rr) \cdot K(\rr ') \right) V (\rr ') \,
\end{equation}
where $\mathcal{K}$ has been parameterized in factorized form in order to exploit the \textit{kernel trick}.

In Eq.~\ref{eq:linear-attn-def}, we define the so-called queries, keys, and values $Q, K, V \in \mathbbm{R}^d$ as learnable local linear transformations of the input field $\phi (\rr) \in \mathbbm{R}^d$:
\begin{equation}
\label{eq:qkv}
    Q(\rr) = W^Q \phi (\rr) \, , \; K(\rr) = W^K \phi (\rr)  \quad \text{and} \quad V(\rr) = W^V \phi (\rr) \; .
\end{equation}

The functional form of Eq.~\ref{eq:linear-attn-def} should be contrasted with an equivalent definition of the standard \texttt{softmax} nonlinear attention~\cite{vaswani_attention_2017, dosovitskiy_image_2020} kernel head
\begin{equation}
\label{eq:attention_head}
    \mathcal{K} _{\text{sm}} (\rr, \rr ') = 
    \frac{ e^{Q (\rr) \cdot K(\rr ') } }{\int \dd[3]{\rr '} e^{Q (\rr) \cdot K(\rr ') } }
\end{equation}
where a similar kernel trick cannot be exploited.

\begin{figure*}[!t]
    \centering
    \begin{minipage}{0.49\textwidth}
        \includegraphics[width=0.6\linewidth]{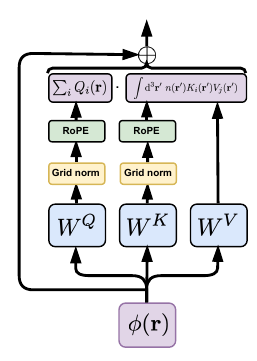}
    \end{minipage}\quad
    \begin{minipage}{0.49\textwidth}
        \includegraphics[width=0.6\linewidth]{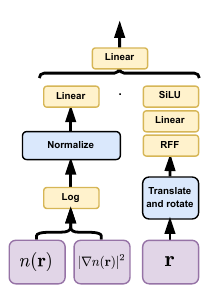}
    \end{minipage}
    \caption{
        Diagrammatic representations of internal connectivities of GDA sub-components.
        \textbf{Left}: The linear attention module enabling the nonlocal behavior of resulting functionals.
        \textbf{Right}: The field embedding module lifting the local field representations to a high-dimentional space.
    }
    \label{fig:layer-diagrams}
\end{figure*}

Large integration grids of $N_g > 10^6$ points are not uncommon in modern DFT calculations. Therefore, the naive \texttt{softmax} self-attention operation between all grid points with asymptotic scaling $\mathcal{O}(N _g ^2)$ is prohibitively numerically expensive. For large point cloud datasets like densities $n(\rr)$ on Becke grids, this is necessary to avoid materializing the full $N_g \times N_g$ kernel matrix $\mathcal{K}(\rr _i, \rr _j)$, incurring prohibitively large memory requirements. Instead, in Eq.~\ref{eq:linear-attn-def}, we first contract $K(\rr)$ and $V(\rr')$, then perform the integral and, finally, multiply by $Q(\rr)$.

Additionally, we apply custom \textit{grid normalization}, to field projections of the input field $\phi$ defined in Eq.~\ref{eq:qkv} before performing the attention operation in Eq.~\ref{eq:linear-attn-def}. Grid normalization amounts to normalizing keys $K$ and queries $Q$ in the following way:
\begin{equation}
\label{eq:grid-norm}
    f(\rr) \; \mapsto \; \frac{f(\rr)}{\sqrt{\int \dd[3]{\rr} n(\rr) (f(\rr))^2}} \, ,
\end{equation}
in an component-wise way for multi-component fields. We note that the choice of leaving values $V$ unnormalized has been called \textit{Fourier} linear attention as opposed to \textit{Galerkin} linear attention where keys and values are normalized.

Furthermore, in order to ensure that the resulting kernel $\mathcal{K}(\rr - \rr ') = Q(\rr) \cdot K(\rr)$ is purely a function of the relative coordinate, we employ rotary positional encoding (RoPE)~\cite{su_roformer_2023, li_transformer_2023}. The RoPE mechanism relies on the composition propery of representations $D$ of the rotation group $SO(2)$ in two dimensions: $D(\alpha) D(\alpha ') = D(\alpha + \alpha ')$ for angles $\alpha$ and $\alpha'$.

We generalize the RoPE mechanism to three spatial dimensions by setting $\alpha = \kk \cdot \rr$ for an input at point $\rr \in \mathbbm{R}^3$ and include $\kk \in \mathbbm{R}^3$ in the learnable parameters of the model. Therefore, we have
\begin{equation}
\label{eq:group-composition}
    D(\kk \cdot \rr) D(-\kk \cdot \rr') = D(\kk \cdot \rr) D ^{-1} (\kk \cdot \rr') = D(\kk \cdot (\rr - \rr')) \; .
\end{equation}
To construct an orthogonal representation $D$ operating on the input field components $\phi _i (\rr)$ and exploit Eq.~\ref{eq:group-composition}, we first note that a trivial representation of SU(2) on $\mathbbm{R}^2$ is given by simple rotation matrices
\begin{equation}
    D_{\mathbbm{R}^2} (\alpha) = \mqty[\cos \alpha & \sin \alpha \\ -\sin \alpha & \cos \alpha] \; .
\end{equation}
Therefore, a multi-component input field $\phi (\rr)$ with an even number of components, we define the 3D RoPE as a trivially reducible orthogonal representation $\phi (\rr) \mapsto D_{\mathbbm{R}^{2 n}} (\kk \cdot \rr) \, \phi (\rr) $ with
\begin{equation}
\label{eq:reducible-rep}
    D_{\mathbbm{R}^{2 n}} (\kk _1 \cdot \rr, \ldots ,\kk _n \cdot \rr) = \mqty[
        \dmat{
            \cos (\kk _1 \cdot \rr) & \sin (\kk _1 \cdot \rr) \\ -\sin (\kk _1 \cdot \rr) & \cos (\kk _1 \cdot \rr),
            \ddots,
            \cos (\kk _n \cdot \rr) & \sin (\kk _n \cdot \rr) \\ -\sin (\kk _n \cdot \rr) & \cos (\kk _n \cdot \rr),
        }
    ]
\end{equation}
depending on $n$ rotation angles defined by trainable parameters $\KK = \{ \kk _1, \ldots ,\kk _n \}$. The block-diagonal structure in Eq.~\ref{eq:reducible-rep} mixes adjacent components of the input field in a way that reflects the target $SO(2)$ multiplication property. In that case, the query-key product in Eq.~\ref{eq:linear-attn-def}
\begin{align*}
    \mathcal{K}(\rr, \rr') =& Q(\rr) \cdot K(\rr ') = \\
    =& [ D(\KK \cdot \rr) W^Q \phi (\rr) ] ^\top [ D(\KK \cdot \rr ') W^K \phi (\rr) ] = \\
    =& \phi (\rr) ^\top (W^Q) ^\top D ^\top (\KK \cdot \rr) D(\KK \cdot \rr ') W^K \phi (\rr) = \\
    =& \phi (\rr) ^\top (W^Q) ^\top D ^\top (- \KK \cdot (\rr - \rr ') ) W^K \phi (\rr) = \\
    =& \mathcal{K}(\rr - \rr') 
\end{align*}
reduces the linear attention integral to a learnable continuous convolution, if we apply the 3D RoPE transformation as a final step just before integration.

Finally, after the attention layer has been updated, the output field $\phi$ is transformed by a component-wise gated multi-layer perceptron~\cite{shazeer_glu_2020}:
\begin{equation}
\label{eq:swiglu}
    \phi (\rr) \; \mapsto \; W_3 \left( (W_1 \phi(\rr) + \vb{b} _1 ) \odot \sigma (W_2 \phi(\rr) + \vb{b} _2 ) \right) + \vb{b}_3
\end{equation}
with \texttt{SiLU} activations~\cite{elfwing_sigmoid-weighted_2017} $\sigma(x)= \frac{x}{1 + e^{-x}}$. Weights $W_i$ and biases $\vb{b}_i$ are included in trainable parameters and $\odot$ indicates element-wise multiplication.

In summary, the steps comprising one GDA \textit{block} are:
\begin{enumerate}
    \item Evaluate raw queries, keys and values by linear projections in Eq.~\ref{eq:qkv}.
    \item Normalize queries and keys using grid normalization in Eq.~\ref{eq:grid-norm}.
    \item Apply the 3D RoPE transformation in Eq.~\ref{eq:reducible-rep} to queries and keys.
    \item Evaluate the final linear attention integral in Eq.~\ref{eq:linear-attn-def} using the kernel trick. Add a skip connection.
    \item Normalize~\cite{ba_layer_2016} and apply the gated MLP in Eq.~\ref{eq:swiglu}. Add a skip connection.
\end{enumerate}
Internal connectivity of sub-layers can be found in Fig.~\ref{fig:layer-diagrams}.

\subsection{Field embedding}
\label{appendix:posenc}

\begin{table}[!t]
\label{tab:hyperparams}
\centering
\begin{tabular}{c|c|c|c|c}
    \textbf{Symbol} & \textbf{Name} & \textbf{Value} & \textbf{Domain} & \textbf{Description} \\
    \hline \hline
    $L$ & Number of blocks & 3 & $\mathbbm{N}$ & \vphantom{\thead{a\\b}} GDA block count \\
    \hline
    $d$ & \thead{Field embedding\\dimension} & 128 & $\mathbbm{N}$ & \thead{Dimension of internal field \\ representations within the model} \\
    \hline
    $\sigma$ & RFF kernel scale & 1 & $\mathbbm{R} _+$ & \thead{Gaussian scale used to \\ initialize the RFF embedding layer} \\
    \hline
    $\eta$ & Learning rate & \thead{Scheduled \\ $2 \times 10^{-4} \rightarrow 5 \times 10^{-5}$} & $\mathbbm{R}_+$ & (R)Adam optimizer learning rate~\cite{kingma_adam:_2015, liu_variance_2019} \\
    \hline
    $B$ & Batch size & 288 & $\mathbbm{N}$ & \thead{Number of molecules used for cost \\ gradient estimation} \\
    \hline
    $N_e$ & Number of epochs & 4000 & $\mathbbm{N}$ & \vphantom{\thead{a\\b}} Number iterations over the entire dataset \\
    \hline
    $\lambda$ & Relative cost weight & 1 & $\mathbbm{R} _+$ & \thead{Constant mutiplier for the potential\\matrix penalty in the main text} \\
    \hline
    $\Tilde{\lambda}$ & Weight decay & 0.01 & $\mathbbm{R} _+$ & \thead{Weight decay regularizer\\for the network parameters~\cite{loshchilov_fixing_2017}} \\
    \hline
    $\alpha$ & Enhancement & 2 & $\mathbbm{R} _+$ & \thead{Relative increase in the number of\\features in the middle MLP layer.} \\
    \hline
    $t_1$ & \thead{Learning rate\\annealing start} & $\nicefrac{N_e}{3}$ & $\mathbbm{N}$ & \thead{Epoch at which the learning\\rate annealing starts (see Fig.~\ref{fig:lr-schedule})} \\
    \hline
    $t_2$ & \thead{Learning rate\\annealing end} & $0.95 \, N_e$ & $\mathbbm{N}$ & \thead{Epoch at which the learning\\rate annealing ends (see Fig.~\ref{fig:lr-schedule})} \\
\end{tabular}
\caption{
    The list of relevant hyperparameter choices used in this work.
}
\end{table}

The field embedding layer is used in the GDA architecture to ensure correct symmetry properties of the learned model. Since the model is explicitly coordinate-dependent, we fix a coordinate system to perform the computation with comparable values of resulting \textit{computational} coordinates. We choose the coordinate system in which the mean value (electronic dipole moment)
\begin{equation}
    \bm{\mu} = \frac{1}{N} \int \dd[3]{\rr} n(\rr) \, \rr
\end{equation}
vanishes and the covariance matrix
\begin{equation}
    \Sigma _{i j} = \frac{1}{N} \int \dd[3]{\rr} n(\rr) \left( r _i - \mu _i  \right) \left( r _j - \mu _j  \right) 
\end{equation}
is diagonal. After fixing translations and rotations, individual coordinates $\rr _i$ are independently embedded into a $d$-dimensional space as
\begin{equation}
\label{eq:rff}
    \rr \mapsto \xi (\rr) = \frac{1}{\sqrt{d}}
    \mqty[ \, \cos \left( K_0 \rr \right) & \parallel & \sin \left( K_0 \rr \right) \, ] ^\top \; ,
\end{equation}
where $\parallel$ denotes vector concatenation and $K_0 \in \mathbbm{R}^{\frac{d}{2} \times 3}$ is a trainable matrix randomly initialized with values sampled from the normal distribution $\mathcal{N} (0, \sigma ^ {-2})$. This \textit{random Fourier feature} (RFF) encoding has effectively been used to encode high-frequency maps of continuous coordinates in kernel learning~\cite{rahimi_random_2007}, and has been shown to measurably increase expressivity of various architectures in geometric deep learning~\cite{mildenhall_nerf_2020, gao_nerf_2022, zheng_rethinking_2021}, including in combination with transformers~\cite{li_learnable_2021}.

The mapping $\rr \mapsto \xi (\rr)$ in Eq.~\ref{eq:rff} is designed to take advantage of the (approximate) kernel trick described in the main text, in a related way to to the RoPE mechanism described in Eq.~\ref{eq:reducible-rep}. However, we emphasize that it is often used outside the kernel context, as a featurization layer (called \textit{positional encoding}) for coordinate-dependent neural network inputs. Suppose that a convolution integral with a kernel $\mathcal{K}$ needs evaluation. Denoting Fourier transforms with hats and employing the convolution theorem in combination with Monte Carlo integration, we have:
\begin{equation}
    \int \dd[3]{\rr'} \, \mathcal{K}(\rr -\rr ') f(\rr ') =
    \int \frac{\dd[3]{\kk}}{(2 \pi) ^3} \, \Hat{\mathcal{K}}(\kk) \, \Hat{f}(\kk) \, e^{i \kk \cdot \rr} \approx
    \frac{1}{(2 \pi) ^3} \times \frac{1}{N_s} \sum _j \Hat{f}(\kk _j) \, e^{i \kk _j \cdot \rr}
\end{equation}
for $N_s$ samples $\kk _j \sim \Hat{\mathcal{K}}$, assuming the the kernel $\mathcal{K}$ is a positive distribution on both sides of the Fourier transform. Therefore, the spread of the distribution in determines the initialization of $K_0$ and it is, in turn, determined by the inverse of the spread of $\mathcal{K}$ in real space, constrained by the uncertainty relation. If we choose to initialize the kernel as a Gaussian $\mathcal{N}$, we have $\kk _j \sim \exp(-\frac{1}{2} \sigma^2 \kk ^2) \propto \mathcal{N} (0, \sigma^{-2})$.

Local density field values are embedded into a $d$-dimensional space as well with learnable linear projections of normalized values. We take local field information in the form of $n(\rr)$ and $\gamma (\rr) = |\nabla n(\rr) |^2$ and construct
\begin{equation}
    \overline{\ln n(\rr)} = \frac{\ln n(\rr) - \ev{\ln n(\rr) } }{ \sqrt{\ev{ (\ln n(\rr) - \ev{\ln n(\rr)})^2 }} }
    \quad \text{and} \quad
    \overline{\ln \gamma(\rr)} = \frac{\ln \gamma(\rr) - \ev{\ln \gamma (\rr) } }{ \sqrt{\ev{ (\ln \gamma(\rr) - \ev{\ln \gamma(\rr)})^2 }} } \; ,
\end{equation}
where we use the density average $\ev{\cdot} = \frac{1}{N} \int \dd[3]{\rr'} n(\rr) (\cdot)$, normalized to unity. In practice, we initialize
\begin{equation}
    \phi(\rr) = W^E \; \mqty[
        \overline{\ln (n(\rr) + \epsilon)} \\
        \overline{\ln (\gamma (\rr) + \epsilon)}
    ] 
\end{equation}
with $\epsilon = 10^{-4}$, instead of bare logarithms for numerical stability and $W^E \in \mathbbm{R}^{d \times 2}$ are trainable parameters represented by trainable unit vectors in the main text. After input fields and coordinates have been independently processed, we combine them using learnable weights $W$ and biases $b$
\begin{equation}
\label{eq:embedding}
    \phi (\rr) \mapsto W' ( \phi (\rr) \odot \sigma (W^\xi \xi (\rr) + b^\xi)) + b' \; .
\end{equation}
to produce final density embeddings. We use $\sigma = \texttt{SiLU}$ as an element-wise nonlinear activation function~\cite{elfwing_sigmoid-weighted_2017}. A list of all relevant hyperparameters used in this work can be found in table~\ref{tab:hyperparams}. A diagram of the overall connectivuty of the embedding layer can be found in the right panel of Fig.~\ref{fig:layer-diagrams}.

\begin{figure*}[!t]
    \centering
    \includegraphics[width=.6\linewidth]{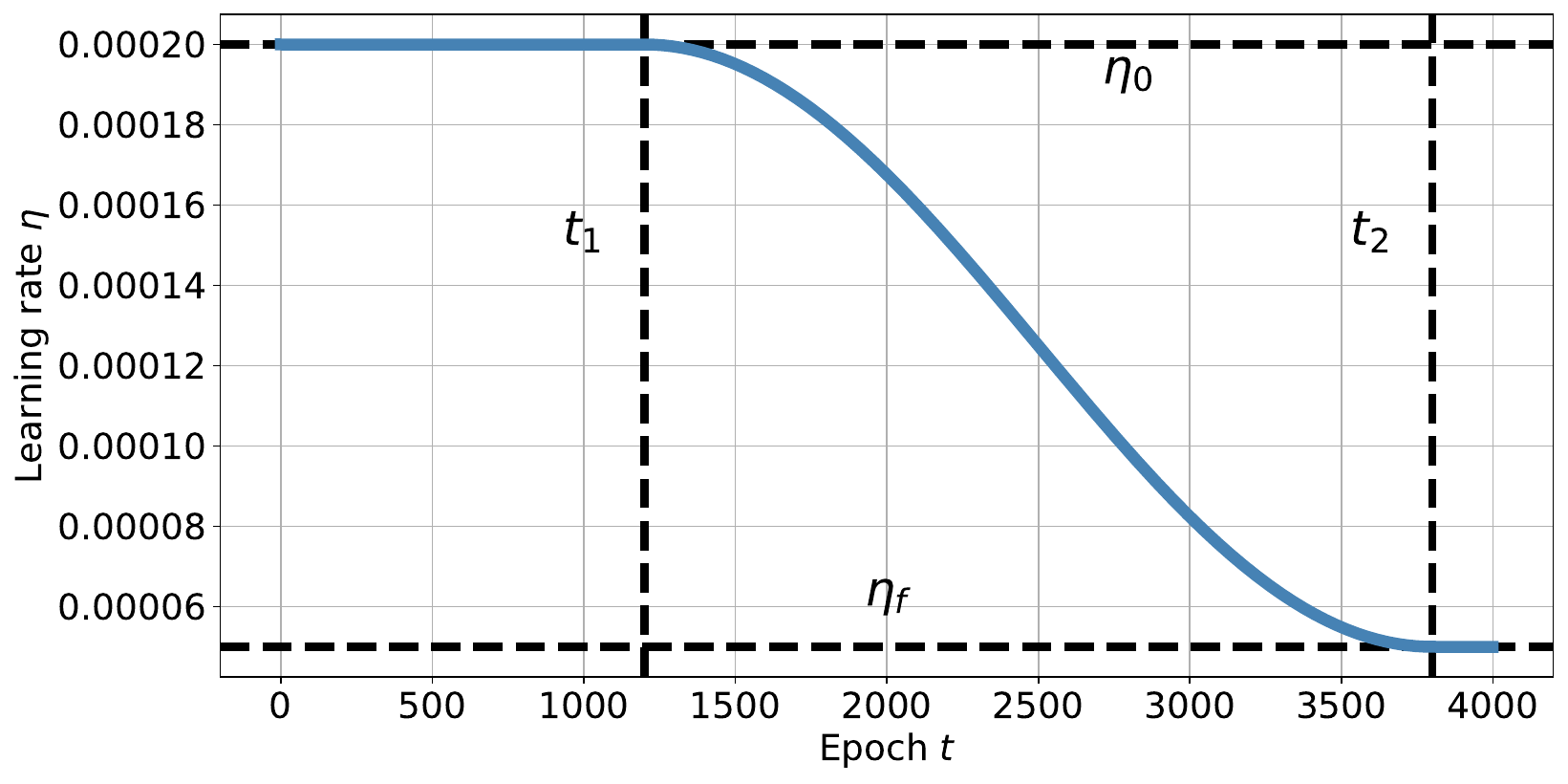}
    \caption{
        The learning rate scheduling used to optimize the full GDA model.
    }
    \label{fig:lr-schedule}
\end{figure*}

\section{Optimization and convergence}

\subsection{Neural-network training}

As noted in the main text, the GDA neural network was trained with $L=3$ blocks, internal field embedding dimension $d=128$, totaling $645 000$ parameters. The rectified Adam~\cite{kingma_adam:_2015} (RAdam)~\cite{liu_variance_2019} optimizer was used for $N_e = 4000$ epochs with learning rate decreasing from $\eta _i = 2 \times 10^{-4}$ to $\eta _f = 5 \times 10^{-5}$ using a customized cosine annealing schedule defined as
\begin{equation}
    \eta _t =
    \begin{cases}
        \eta _i, & \text{if}\ t < t_1 \\
        \eta _f + (\eta _i - \eta _f) \cos ^2 \left( \frac{\pi}{2} \frac{t - t_1}{t_2 - t_1} \right), & \text{if}\ t_1 \leq t \leq t_2 \\
        \eta _f, & \text{if}\ t > t_2 \\
    \end{cases}
\end{equation}
for epoch $t$. The schedule is shown in Fig.~\ref{fig:lr-schedule}. We find that decreasing the learning rate during the course of training helps fine-tune the model in the later epochs when the loss landscape changes at smaller scales. We set $t_1 = N_e / 3$ and $t_2 = 0.95 N_e$.

\subsection{Self-consistent field optimization from first principles}

We use PySCF~\cite{sun_pyscf_2018, sun_recent_2020} for first-principles optimizations of molecular densities, where the default method is the standard SCF Pulay mixing or direct inversion in the iterative subspace (DIIS)~\cite{pulay_convergence_1980}. Using that method, we optimize a random molecule from the training dataset a total of six times -- three times with library~\cite{marques_libxc_2012} versions of targeted meta-GGA functionals~\cite{sun_strongly_2015, furness_accurate_2020, tao_climbing_2003} and three times with the GDA substitution $\tau \rightarrow \tau _\theta$. Results can be seen in the left panel of Fig.~\ref{fig:performance}.

We note that r${}^2$SCAN and TPSS converge in a similar number of mixing iterations to the original functionals, while SCAN (included for comparison) appears to be more sensitive to numerical errors associated with the GDA approximation. We report that this trend persists in most of the other molecules that we examined more closely.

\section{Numerical performance benchmarks}

\begin{figure*}[!t]
    \centering
    \begin{minipage}{0.45\textwidth}
        \includegraphics[width=\linewidth]{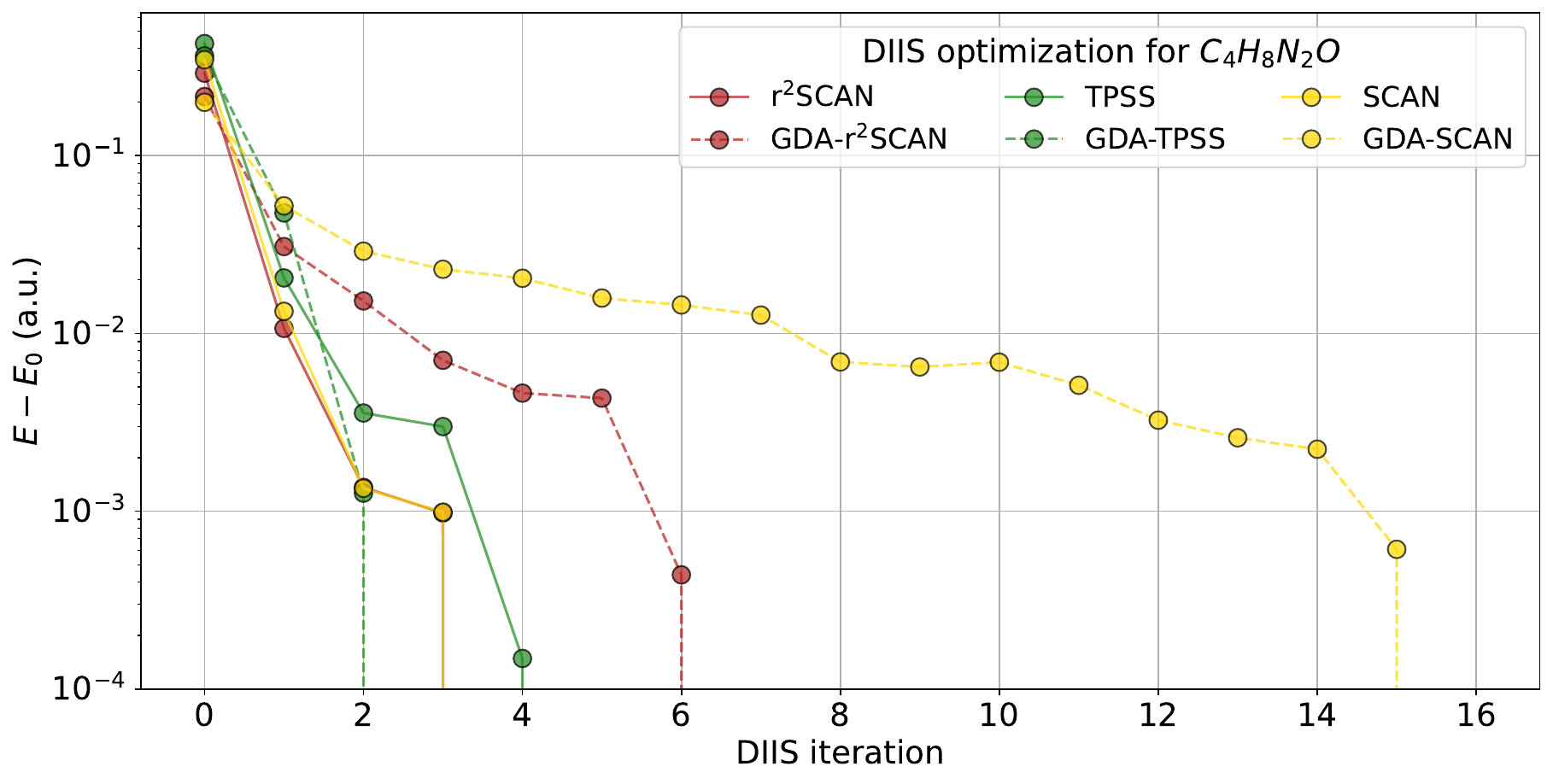}
    \end{minipage}\quad
    \begin{minipage}{0.53\textwidth}
        \includegraphics[width=0.9\linewidth]{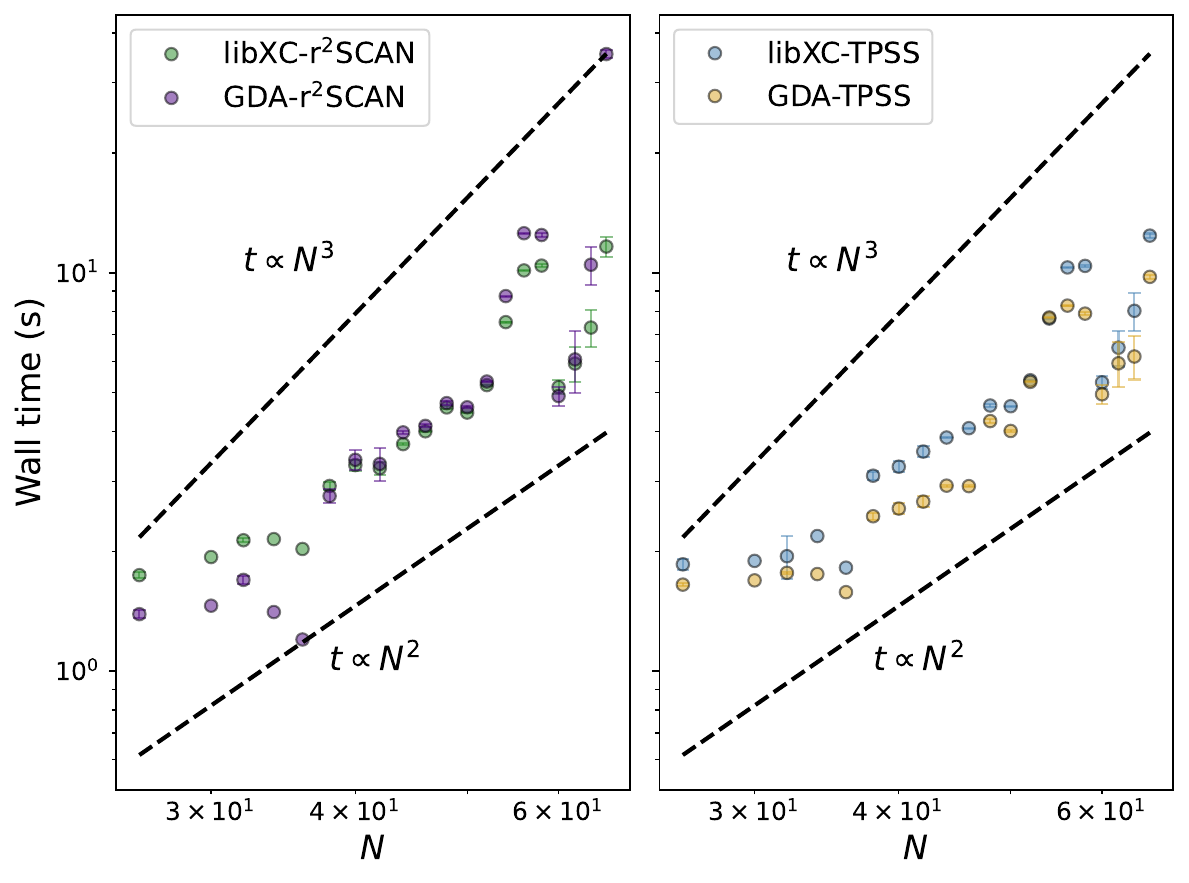}
    \end{minipage}\quad \hfill
    \begin{minipage}{0.49\textwidth}
        \includegraphics[width=\linewidth]{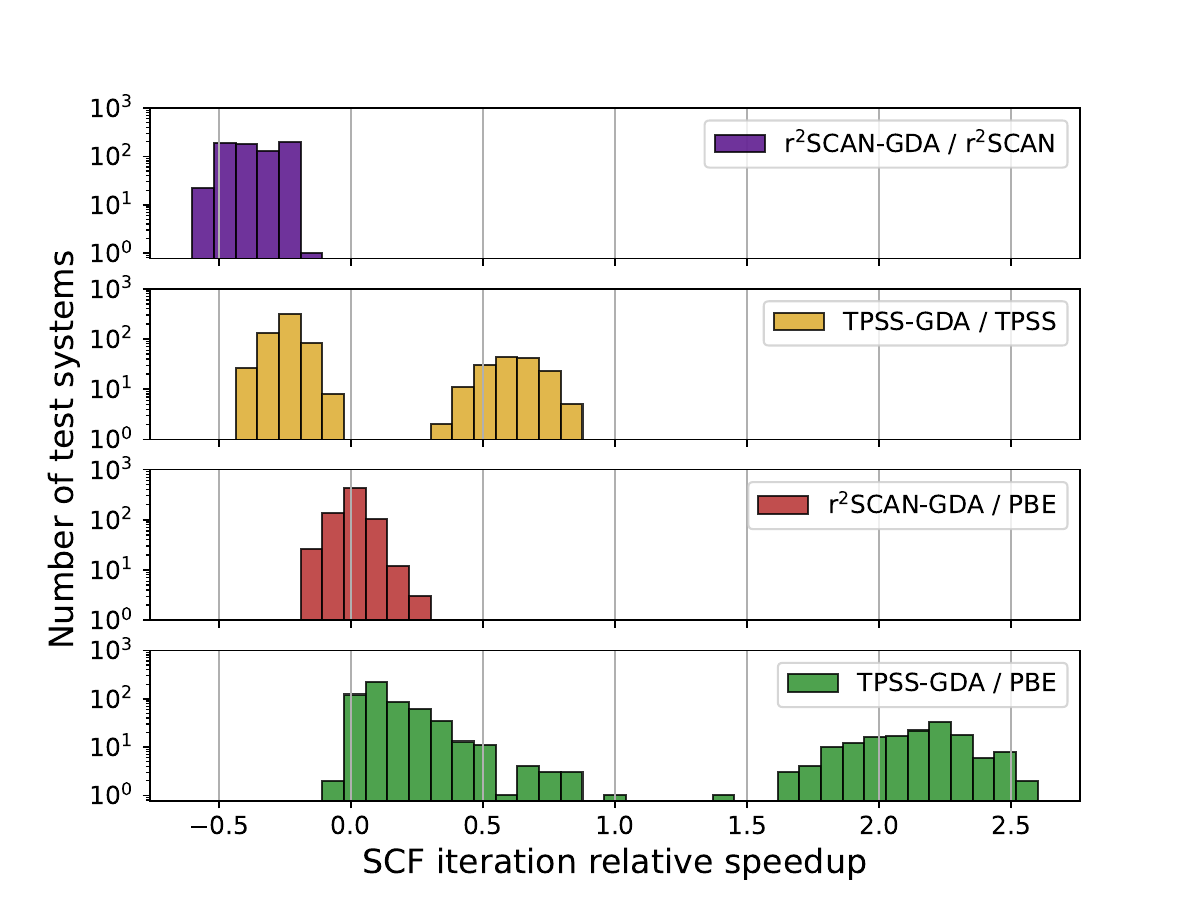}
    \end{minipage}\quad
    \begin{minipage}{0.48\textwidth}
        \includegraphics[width=\linewidth]{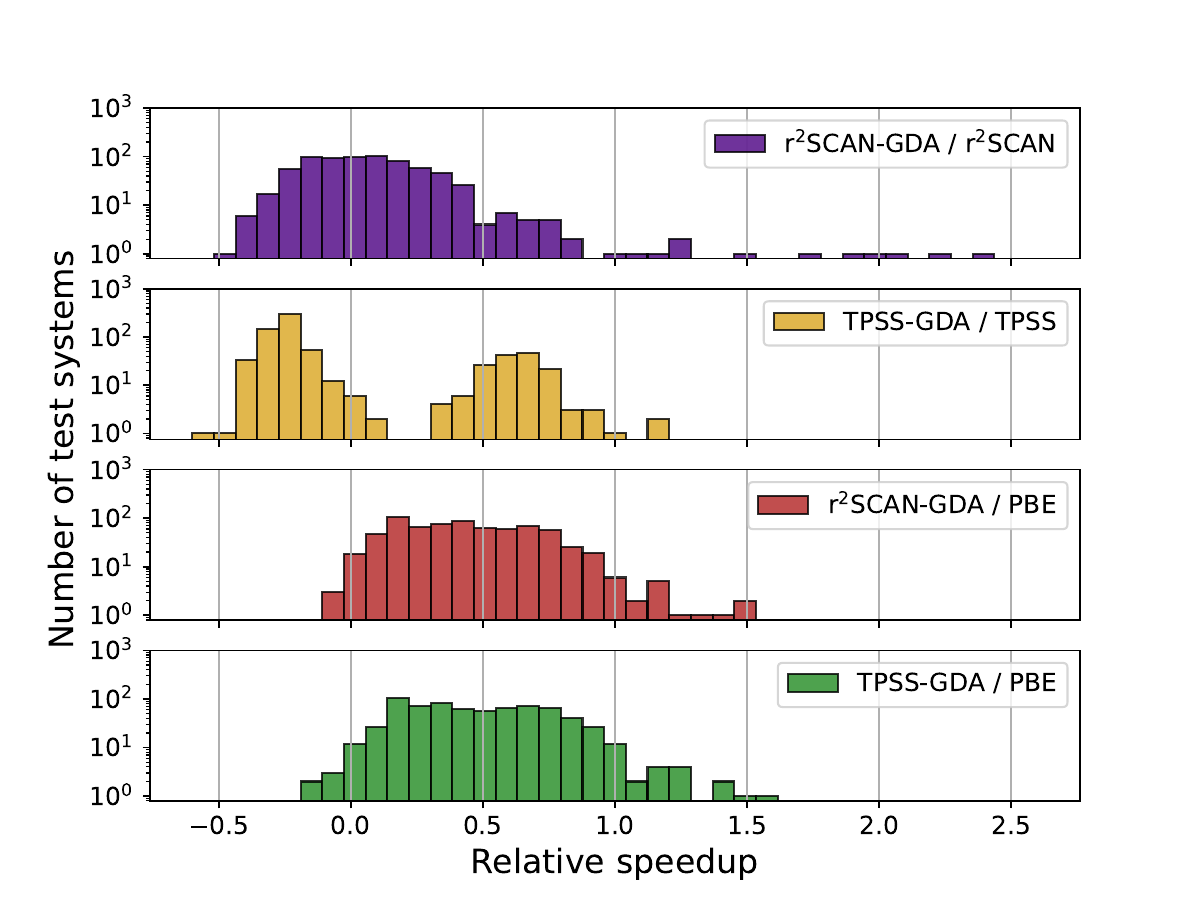}
    \end{minipage}
    \caption{
        Convergence and numerical performance properties of the GDA functionals.
        \textbf{Top left}:
            Convergence properties of the DIIS~\cite{pulay_convergence_1980} optimization used in PySCF~\cite{sun_pyscf_2018, sun_recent_2020}, on an example of a random molecule from the test dataset.
        \textbf{Top right}:
            Total wall clock time required for SCF convergence as a function of the number of electrons $N$. We observe scaling consistent with the expected $\mathcal{O}(N^3)$ in the large-$N$ limit.
        \textbf{Bottom left}:
            Relative speedups defined in Eq.~\ref{eq:relative-speedup} for individual SCF iterations. We observe that the GDA approximation offers up to $50\%$ faster SCF iterations while still not being as efficient as the representative pure-density functional (PBE). 
        \textbf{Bottom right}:
            Relative speedups for total wall clock times. We see that, within our hardware constraints, the GDA approximation offers minor speedups at best. We note that these results are extremely hardware and software dependent. Additionally, we expect that the more fine-tuned models trained on specialized data sets will improve in performance as well as benefit from more optimizations in the GDA code under development.
    }
    \label{fig:performance}
\end{figure*}

In this subsection, we state some numerical heuristics when comparing the efficiency of the GDA approximation when compared to the parent meta-GGA functional. We note that, like many computational benchmarks, it is difficult to control all variables contributing to practical wall-clock times. That is especially true for the case of DFT within the context of parametrized models. Such modern AI models are accelerated using graphical processing units (GPUs), offering speedups coming massive parallelism for some operations (e.g. matrix multiplication).

Therefore, any fair comparison between efficient implementations of \textit{traditional} and machine-learned density and orbital functionals will almost certainly be \textit{apples-to-oranges} to some degree. Results change with different (and rapidly evolving) hardware, suffer from CPU-GPU communication overhead, just to name a few factors. Additionally, desired convergence accuracy and target precision may play an important role in determining the outcomes of such comparisons because modern GPU hardware tends to be heavily optimized for single-precision floating point operations. 

Our testing setup reflects a \textit{workstation} setup with an AMD Ryzen Threadripper 7970X 32-core CPU, Nvidia RTX 4500 (Ada Generation) GPU. Numerical run-time data was collected from independent calculations on 10\% of the total QM7~\cite{blum_970_2009, rupp_fast_2012} dataset discussed in the main text. Heuristics on relative runtime speedups, data on SCF convergence, and individual iteration efficiency can be found in Fig.~\ref{fig:performance}.

To quantify the performance difference, we define the \textit{relative speedup} as
\begin{equation}
\label{eq:relative-speedup}
    r = \frac{t_\text{X} - t_\text{ref}}{t_\text{ref}}
\end{equation}
for a DFT calculation taking $t_X$ time using the GDA functional X, compared against a reference calculation $t_\text{ref}$. In Fig.~\ref{fig:performance}, we compare the performance of GDA approximations against target meta-GGAs as well as PBE~\cite{perdew_generalized_1996} as a representative of the pure-density functional (GGA) family.

Within the scope of our tests, we observe that GDA models are more efficient than their parent functionals on average, based on timings of individual SCF iterations. In the case of TPSS, we observe a more consistent performance boost (up to $50\%$), in addition to the results presented in the main text. In terms of total (wall clock) times, we see that GDA approximations offer moderate speedups for the QM7 training set by requiring a few extra SCF iterations than reference functionals. Heuristically, we observe that this can be controlled by changing the gradient penalty term coupling $\lambda$ in the cost function presented in the main text. We expect that the transferability of GDA models allows fine-tuning on more specific datasets where additional performance gains can be accessed. 

Within the available hardware constraints, we observe that the overall performance gain of GDA approximation puts them somewhere between traditional meta-GGA and GGA functionals, with overall scaling still consistent with the expected asymptotic $\mathcal{O}(N^3)$.

\section{Density functional theory}
\label{appendix:dft}

\subsection{Kohn-Sham density functional theory}

As set up in the main text, we consider isolated molecular systems. A DFT calculation~\cite{martin_electronic_2020} outputs an approximation to the ground state density $n_0$ minimizing the total energy: $n_0 (\vb{r}) = \argmin _{n} E[n]$ where
\begin{equation}
\label{eq:total_energy}
    E[n] = T[n] + E_\text{ext}[n] + E_H[n] + E_\xc[n]
\end{equation}

Known terms in the total energy functional $E[n]$ given in Eq.~\ref{eq:total_energy} are the external contribution $E_\text{ext}[n]$
\begin{equation}
    E_\text{ext}[n] = \int \dd[3]{\rr} n (\rr) \, \vext (\rr)
\end{equation}
capturing the effects of the atomic Coulomb interaction $v _\text{ext} (\rr)$ and the direct Hartree component capturing the classical electronic Coulomb interaction:
\begin{equation}
    E_\text{H}[n] = \frac{1}{2}\int \dd[3]{\rr} \int \dd[3]{\rr '} \frac{n (\rr) n (\rr ')}{\left| \rr - \rr ' \right|} \; .
\end{equation}

Kohn-Sham DFT~\cite{kohn_self-consistent_1965} framework assumes that the density $n_0$ comes from an effective system of non-interacting electrons with orbitals $\Psi = \{ \psi _k (\rr) \, | \, k = 1, \ldots , N \}$ where $N$ is the number of electrons in the system. In that case, the kinetic term in Eq.~\ref{eq:total_energy} evaluates to
\begin{equation}
    T[n] \; \mapsto \; T _{\text{KS}} [n] = \sum _a n_a \int \dd[3]{\rr} \; \psi ^* _a (\rr) \left( -\frac{1}{2} \nabla ^2 \right) \psi _a (\rr) =
    \frac{1}{2} \sum_a n_a \int \dd[3]{\rr} \left| \nabla \psi _a (\rr) \right| ^2 \; .
\end{equation}

After the desired kinetic and XC functional has been specified, the constrained minimization of the total energy functional given in Eq.~\ref{eq:total_energy} can proceed, enforcing orbital normalization with Lagrange multipliers $\epsilon _a$:
\begin{equation}
\label{eq:ks_lagrangian}
    \mathcal{L} [\Psi ] = E[\Psi ] + \sum _a \epsilon _a \int \dd[3]{\rr} \left| \psi _a (\rr) \right| ^2 = 
    T_\text{KS} [\Psi ] + U[\Psi] + \sum _a \epsilon _a \int \dd[3]{\rr} \left| \psi _a (\rr) \right| ^2
\end{equation}
where we choose to trivially rewrite all density functionals as orbital functionals and collect all of the different kinds of potential energy into $ U [\Psi] = E_\text{ext} [\Psi ] + E_\text{H} [\Psi ] + E_\xc [\Psi ]$. Trivial functional differentiation yields:
\begin{equation}
\label{eq:ks_equation}
    \fdv{\mathcal{L} [\Psi] }{\psi ^* _a (\rr) } = \left( - \frac{1}{2} \laplacian + v_\eff [n] (\rr) - \epsilon _a \right) \psi _a (\rr )
    \quad \Longrightarrow \quad
    \left(- \frac{1}{2} \laplacian + v_\eff [n] (\rr) \right) \psi _a (\rr) = \epsilon _a \, \psi _a (\rr) \, ,
\end{equation}
known as the Kohn-Sham (KS) equation. Mathematically speaking, the KS Eq.~\ref{eq:ks_equation} is a non-linear partial differential equation for the unknown orbitals $\Psi$. The source of non-linearity, other than the classical Coulomb term, comes from exchange and correlation effects built into the approximate functional $E _\xc$ that gives rise to the effective one-electron potential
\begin{equation}
\label{eq:veff}
    v_\eff (\rr) = \fdv{U [n ]}{n (\rr)} = \vext (\rr) + \int \dd[3]{\rr '} \frac{n (\rr ')}{\left| \rr - \rr ' \right|} + \fdv{E _\xc [n]}{n (\rr)} \, .
\end{equation}

The last term in Eq.~\ref{eq:veff} is sometimes labeled as the XC potential $v _\xc (\rr)$ and in it highlights the importance of being able to calculate functional derivatives of the XC functional efficiently.

Numerically, Eq.~\ref{eq:ks_equation} is usually solved in a self-consistent (SCF) manner, ensuring that $n (\rr) = \sum _a n _a | \psi _a (\rr) | ^2$. The SCF procedure takes the form of fixed-point iteration of the following two steps, starting from the initial guess for $n (\rr)$:
\begin{itemize}
    \item Update orbital estimates $\Psi$ by solving Eq.~\ref{eq:ks_equation} for a fixed density.
    \item Update density estimate through $n (\rr ) = \sum _a n _a | \psi _a (\rr) | ^2$. 
\end{itemize}
The loop is terminated after successive density and energy estimates stop changing beyond a given tolerance.

\subsection{Calculations in the basis of atomic orbitals}

For isolated molecular systems, Gaussian basis sets are a common approach of representing atomic orbitals used in density expansions. In this subsection, we derive all of the expressions used during training to evaluate the GDA effective one-particle Hamiltonian
\begin{equation}
\label{eq:effective-hamiltonian}
    H_\eff = \frac{\vb{p}^2}{2} + v_\eff [n] (\rr) 
\end{equation}
from the neural-network functional using automatic differentiation, with $\vb{p}$ being the one-particle momentum operator. With the definition of Eq.~\ref{eq:effective-hamiltonian}, the KS equations simply read $H_\eff \ket{\psi_a} = \epsilon _a \ket{\psi _a}$.

In the following, we use Greek indices $\{ \mu , \nu, \ldots \}$ for the fixed atomic orbital (AO) basis and latin $\{ a , b, \ldots \}$ indices for the molecular orbital (MO) basis. The two bases are related by the LCAO (linear combination of atomic orbitals) coefficients $C$ as $\ket{\psi _a} = \sum _\mu C _{\mu a} \ket{\chi _\mu}$. The density matrix $\Gamma$ is usually defined in the AO basis through
\begin{equation}
    n (\rr) = \sum _{\mu \nu} \Gamma _{\mu \nu} \chi _\mu (\rr) \chi _\nu (\rr) \; .
\end{equation}

The full effective Hamiltonian matrix in the AO basis decomposes as:
\begin{equation}
    F_{\mu \nu} = \bra{\chi _\mu} H_\eff \ket{\chi _\nu} = \pdv{E}{\Gamma _{\mu \nu}} = T _{\mu \nu} + V _{\mu \nu} + J _{\mu \nu} + X _{\mu \nu} \; ,
\end{equation}
where $T$, $V$, $J$, and $X$ are kinetic, nuclear, Coulomb and XC matrices, respectively. These quantities are readily available in quantum chemistry software packages and we save $T$ as a part of the training dataset. The Kohn-Sham kinetic energy functional is of special interest in this work. Employing partial integration, we get
\begin{align*}
    T[n] =& \sum _a n_a \bra{\psi _a} \frac{\vb{p}^2}{2} \ket{\psi _a} = \\
    =& \sum _a n_a \int \dd[3]{\rr} \psi ^* _a (\rr) \left( - \frac{1}{2} \nabla ^2 \right) \psi _a (\rr) = \frac{1}{2} \sum _a n_a \int \dd[3]{\rr} \left| \nabla \psi _a (\rr) \right| ^2 = \\
    =& \sum _{\mu \nu} \left( \sum _a n_a C_{\mu a} C_{\nu a} \right) \times \frac{1}{2} \int \dd[3]{\rr} \, \nabla \chi _\mu (\rr) \cdot \nabla \chi _\nu (\rr) = \sum _{\mu \nu} \Gamma _{\mu \nu} T _{\mu \nu} = \Tr \left( \Gamma T \right) \; ,
\end{align*}
where we identify the kinetic energy matrix $T_{\mu \nu}$ that goes into the cost function defined in the main text. Other similar matrices in the AO basis work out to be:
\begin{equation}
\begin{gathered}
    T _{\mu \nu} = \frac{1}{2} \int \dd[3]{\rr} \, \nabla \chi _\mu (\rr) \cdot \nabla \chi _\nu (\rr) \; , \quad
    V _{\mu \nu} = \int \dd[3]{\rr} \, v _{\text{ext}} \, (\rr) \chi _\mu (\rr) \chi _\nu (\rr) \quad \text{and} \quad \\
    J _{\mu \nu} = \sum _{\alpha \beta} \Gamma _{\alpha \beta} \int \dd[3]{\rr} \int \dd[3]{\rr '} \frac{\chi _\mu (\rr) \chi _\nu (\rr) \chi _\alpha (\rr') \chi _\beta (\rr')}{\left|\rr - \rr ' \right|} \; .
\end{gathered}
\end{equation}

The XC matrix $X$ is the only component we need to evaluate using from the GDA functional in order to regularize the training loop. We only need to evaluate the derivative of the total XC energy with respect to the input density matrix because:
\begin{equation}
\label{eq:xc-mat}
    \hat{X} _{\mu \nu} = \pdv{\hat{E} _\xc}{\Gamma _{\mu \nu}} = \int \dd[3]{\rr} \, \fdv{\hat{E} _\xc}{n (\rr)} \pdv{n (\rr)}{\Gamma _{\mu \nu}} = \int \dd[3]{\rr} \, \hat{v} _\xc (\rr) \, \chi _\mu (\rr) \chi _\nu (\rr) \; .
\end{equation}
where we put hats on quantities estimated using the GDA functional approximation. These derivatives can be calculated efficiently using automatic differentiation at the cost of evaluating the functional itself.

\bibliography{references}